# Modelling nuisance parameters in linear models: Formal and empirical comparisons between the two-way Anova and Ancova.

Carlos A. Martínez[a]*, Nelson A. Cruz[b], Sandra E. Melo[c], Oscar O. Melo[b]

*[a]Universidad Nacional de Colombia. Sede Bogotá. Departamento de Producción Animal., Carrera 45 # 26-85, Bogotá D.C., Colombia, ORCID:* 0000-0003-2842-5580. *[b]Universitat de les Illes Balears, Facultad de Ciencias Matemáticas e Informática, Cra. de Valldemossa, km 7,5. 07122 Palma (Illes Balears), ORCID:* 0000-0002-7370-5111. *[c]Universidad Nacional de Colombia, Sede Bogotá. Departamento de Agronomía, Carrera 45 # 26-85, Bogotá D.C., Colombia, ORCID:* 0000-0002-4875-7657 *[d]Universidad Nacional de Colombia. Sede Bogotá. Departamento de Estadística, Carrera 45 # 26-85, Bogotá D.C., Colombia, ORCID:* 0000-0002-0296-4511

*Corresponding author. Email: camartinezn@unal.edu.co
Ciudad Universitaria, Carrera 45 # 26 - 85 Edificio Uriel Gutiérrez
Bogotá D.C.; Bogotá D.C; Postal code: 111321

## Data availability statement.

*Piglet experiment environmental data.*

The portion of these data corresponding to covariates (environmental data) along with the diet, and block, are openly available in Zenodo.org at https://zenodo.org/records/22655402, reference number 22655402. The complete dataset that supports the findings of this study are available from Universidad Nacional de Colombia. Restrictions apply to the availability of these data, which were used under license for this study. Data are available from the authors with the permission of Universidad Nacional de Colombia.

*Radish experiment data*.

The radish experiment data are openly available in Zenodo.org at https://zenodo.org/records/22655402, reference number 22655402.

## Conflict of interest disclosure.

There are not financial/commercial conflicts of interests.

## Acknowledgements.

This study was supported by the Colombian government through Ministerio de Agricultura y Desarrollo Rural – MADR and AGROSAVIA.

## Abstract

The analysis of experimental or observational data often requires controlling for the effects of variables of secondary interest. Two common approaches to account for such parameters are grouping experimental or observational units into homogeneous groups (blocking), and observing quantitative variables on each one of them (covariance analysis). Often, both can be used and; therefore, comparing their performance poses a research question. Specifically, under randomized complete block designs where the blocks are defined by quantitative variables, there is interest in comparing the two-way Anova (TWA) with analysis of covariance (Ancova) models. Motivated by real-life problems coming from statistical consulting, we compared the performance of linear models considering block effects and those considering covariate effects from analytical and empirical perspectives. Models assumed Gaussian errors and included the effects of T treatments. The formal comparison relied on the theory of orthogonal projections and vector spaces under containment relationships of column spaces of the design matrices induced by conditions on the experimental design or data structure. We found asymptotic and small-sample conditions for the TWA to be at least as precise as the Ancova, as well as for these models to be equivalent. As to the empirical perspective, models were compared through a simulation study motivated by agronomical experiments. The results suggested that when the covariates show complex heterogeneity patterns, the Ancova outperforms the TWA with blocks defined by traditional approaches in terms of precision of location parameters, but not in hypothesis testing accuracy.

## 1. Introduction

A relevant feature of data analysis in different fields is the necessity of controlling for the effects of variables of secondary interest that may impact the response variable; thus, these are considered nuisance parameters. In experimental design, this is known as local control and it is expected to produce more precise comparisons among treatment means (Kuehl, 2000; Melo, López and Melo, 2020). Two approaches for local control are: 1) clustering experimental or observational units into well-differentiated and homogeneous groups called blocks (blocking) and 2) observing the values of a set of $p$ quantitative variables called covariates or concomitant variables on each experimental/observational unit resulting in the so-called Ancova model. Thus, under the blocking approach, the statistical model includes the effects of the levels of a qualitative variable, while under the Ancova model, it considers the regression coefficients of each covariate.

Furthermore, block definition may be immediate when the levels of the blocking factor occur naturally, for example, breed in an animal science experiment, or gender in a clinical trial. Alternatively, the original scale of the variable inducing the heterogeneity to control for, may be numerical (typically continuous), like age, height, temperature or weight. In this setting, the block definition does not follow naturally and there is uncertainty about the number of blocks and the thresholds defining them; moreover, this task becomes far more complex when there are several covariates. Expert judgement can be employed to perform an ad hoc block definition, alternatively, data-driven approaches like cluster analysis may be used. A related problem comes from agronomical studies, where soil heterogeneity is pervasive, thus, this is a factor to control for in most experiments. This phenomenon comes from the high variability that soil exhibits due to topography (Florinsky, 2016), chemical, physical, and biological processes (Behera et al., 2018). The most frequent approach to local control of soil heterogeneity is blocking, which implies splitting the field into $B$ plots (blocks) (Schwenke, 1997; Yang and Juskiw, 2011). Ideally, blocking is based on data from soil analyses of the experimental field; notwithstanding, many soil properties of agronomical relevance may exhibit very complex variation patterns; for example, chemical composition variables (Goovaerts, 1998; Behera et al., 2018). In these cases, especially when the number of soil variables is high, blocking may become almost impossible; on the other hand, precision agriculture yields data that can be used as covariates to control for soil heterogeneity.

The randomized complete blocks design is one of the most used experimental designs to perform local control, and for Gaussian responses, the two-way Anova (with or without interaction) and the Ancova models are frequently used to analyze data from this design in several disciplines. When fitting these models, the main interest is on inferring estimable functions (as defined in linear models theory) of treatment means. Most studies comparing these models are empirical (Wu and McLean, 1994; Schewenke, 1997; Klockars, Potter and Beretvas, 1999; Klockars and Beretvas, 2001; Yang and Juskiw, 2011), in contrast, very few attempts from a formal perspective are available. Cox (1957) and Feldt (1958) used analytical approximations to compare the two-way Anova and the Ancova models that were focussed on the precision of pairwise differences of treatment means and considered a single covariate. However, analytical comparisons that hold for an arbitrary number of covariates and any estimable function are not available and, consequently, there is a need for such studies in order to draw conclusions based on formal results, at least under particular settings. Recently, three statistical consultancies posed an interesting problem regarding the Ancova and two-way Anova models under particular data structures; this consulting problems motivated the theoretical developments presented in this study.

This paper is focused on linear fixed effects models with independent and identically distributed Gaussian errors considering the effects of $T$ groups (the effects of primary interest) and performing local control in two different ways: by incorporating the effects of $B$ blocks (two-way Anova) or the regression coefficients of a collection of $p$ numerical variables (Ancova). The objective was to compare the performance of these statistical models using analytical and empirical approximations, focusing on different scenarios induced by real-life problems in animal sciences and agronomy.

## 2. Methods

This part is organized as follows. Section 2.1 presents the motivating problems. Then, the models to be compared along with some relevant notation, are introduced in section 2.2, while 2.3 presents the approach to perform an analytical model comparison and our main theoretical results. The remaining subsections describe data analyses.

### *2.1 Motivating problems*

We focus on five real-life problems coming from animal sciences and agronomy, namely, local control for soil heterogeneity in agronomical studies under complex spatial variability patterns, evaluating the impact of salinity on plant development, and the design of three experiments in commercial pork and layer production systems.

#### *2.1.1 Controlling for soil heterogeneity in agronomical studies.*

Precision agriculture, a management strategy based on the analysis of site-specific spatial and temporal data (Vanden Heuvel, 1996; Zhang, Wang and Wang, 2002), is being used more and more often. In particular, it offers an alternative to conventional blocking strategies that permits either, a better definition of blocks, if possible, or an alternative analysis scheme that uses observations from $p$ site-specific soil properties as covariates.

The most frequent approach to local control of soil heterogeneity is blocking, which in this particular context implies splitting the field into B plots (blocks) in such a way that soil properties are homogeneous within block and heterogeneous between blocks. Thus, the efficiency of blocking depends on the capacity to create well-differentiated and homogeneous groups of experimental units. The absence of soil composition records to define the blocks is frequent, so blocking is performed *ad hoc* by looking at very general terrain characteristics, like inclination or the presence of bodies of water inducing humidity gradients, and then splitting it into approximately equally sized plots. Thus, under the presence of complex soil heterogeneity patterns, it is very likely that the usual procedures to blocking end up in loss of efficiency (smaller precision). Specifically, the research question posed here is: At what extent accuracy and precision are undermined by the traditional blocking strategy under the presence of complex soil heterogeneity patterns?

*2.1.2 Three statistical consultancies and an agronomical experiment leading to the same theoretical problem.*

The three consultancies came from the field of animal sciences and were asked by independent researchers. The first one stemmed from a pork production system where lactating sows are kept under a controlled environment. The system uses industrial air extractors located on one side of the room, that force air to cross by a column of fins with cold water flowing down, which is located on the other side of the room. Sows are placed in rectangular farrowing pens arranged side by side forming several rows across the farrowing room as shown in Figure 1; notice that these rows are oriented perpendicularly to air flow. The sows and their progeny remain there for the whole lactation period. The consultant wanted to apply different treatments with each farrowing pen as experimental unit, but she was concerned with the row effect, because even though the system provides real-time temperature values for the whole room, preliminary temperature and relative humidity measurements showed changes across rows, but not, within rows (Figure 1).

Thus, they were planning to locate sensors to measure these environmental variables at each row (one sensor per row). This kind of trials consider the average minimum and maximum temperature and relative humidity during the experimental period. Another potential covariate was the distance from the side of the room where air comes from, to the front of each row of farrowing pens. Notice that, in this experiment, covariates take the same value within each row (block), this is the key property featured by this problem.

The second consultancy was quite similar and came from a poultry farming operation, specifically, egg production. Groups of three layers are kept in cages positioned side by side, forming rows arranged in three levels, one beneath the other, within the same shed; also, each shed had two sides (Supplementary Figure 1). Besides temperature and relative humidity, they measure luminous flux, amount of carbonic dioxide, ammonia, wind speed and some environmental organic acids. Researchers were planning a complete blocks design with cage as experimental unit, and blocks corresponding to groups of adjacent cages within the same row as blocks. Furthermore, due to logistic and pragmatic considerations, a measurement instrument would be assigned to a set of adjacent cages corresponding to each block-treatment combination. Consequently, in this experiment, the value of the covariates for experimental units pertaining to the same treatment-block subclass, would have the same value. Supplementary Figure 1 portrays the experimental arrangement for this experiment.

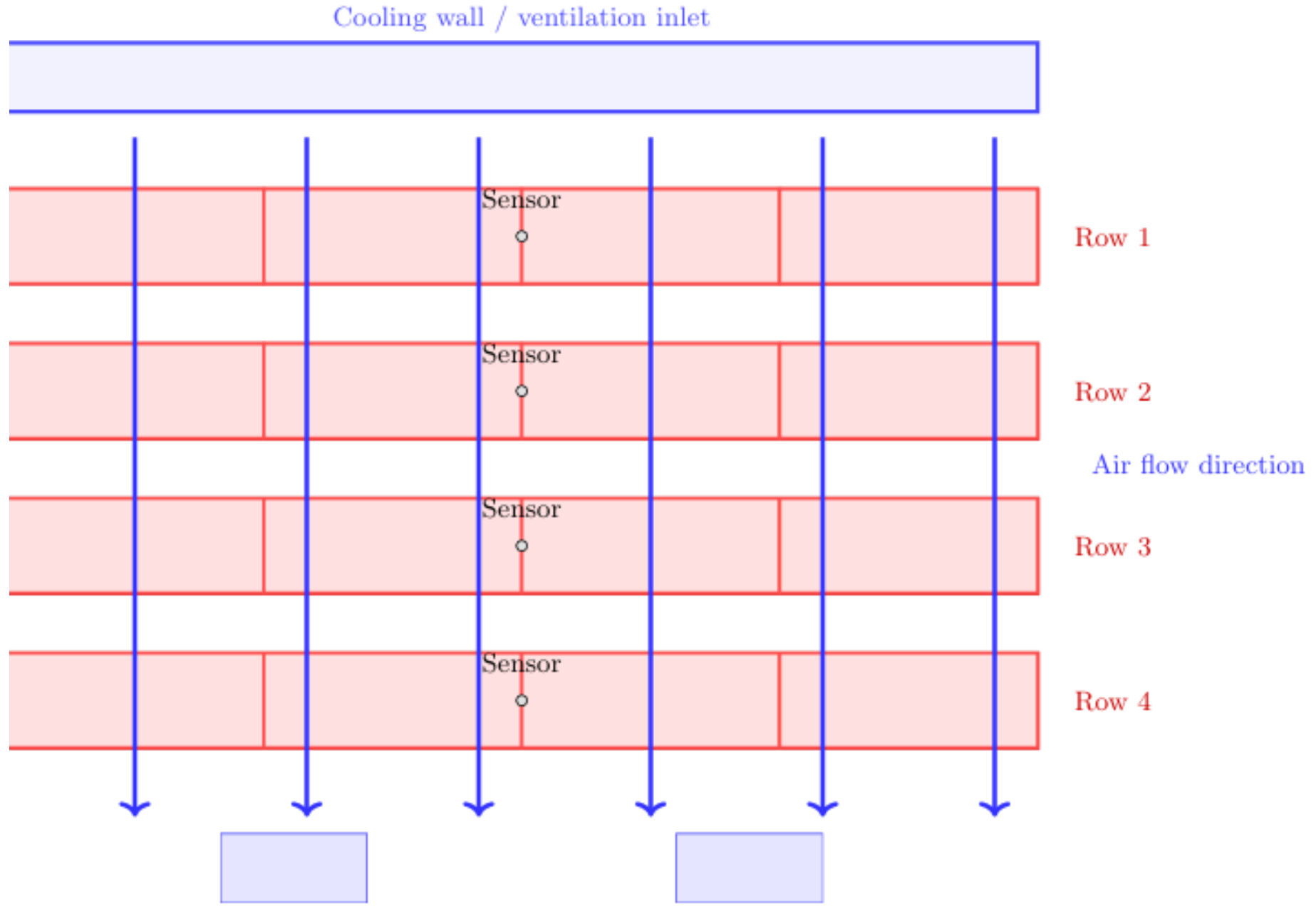

**Figure 1**. Sketch of the lactating sows experiment. Each row corresponds to a block. Covariates such as temperature or humidity have the same value within each row.

Finally, the third problem emerges from the pork industry as well, but this time animals are not kept in a controlled environment. The experiment involved the application of five diets to growing piglets located at the Universidad Nacional de Colombia's research centre Marengo, situated in the municipality of Mosquera, department of Cundinamarca, Colombia. The herd is small and it could not house the total number of experimental units (pens with four piglets) that was planned; thus, three batches of piglets were used. Once a batch finished the 49 days experimental period, the pens were disinfected and the next batch was placed to start a new cycle. Notice that experimental units were not the same and that all treatments were randomized within each batch; thus, it was a randomized complete blocks design with batch as block (three blocks). Marengo has a private weather station and the researchers were interested in using records collected by this station in the data analysis. Therefore, several meteorological variables of interest, averaged over the period occupied by each batch, could be used as covariates. Thus, there was interest in addressing the impact of the inclusion of information from environmental covariates. Notice that this problem leads to exactly the same data structure discussed for the other pork experiment. As to the first two scenarios there is no available data because we do not have access to records from the first one (lactating sows), and the second experiment has not been carried out yet. Fortunately, we had access to records from the third one; in this paper we focus on two response variables: average body weight at 70 days (BW70; kg) and average daily weight gain at 70 days (ADG70; $kg \cdot d^{-1}$).

The three problems have a particular and quite interesting restriction, the value of each covariate remains constant between each block or treatment-block subclass. The applied researcher is interested in addressing the impact of either, replacing the block effects by the effects of the covariates in the model, or adding the covariates to the two-way Anova model. In turn, it raises interesting questions from a theoretical perspective regarding the impact of this pattern on the estimability and precision of linear functions of treatment means. At first glance, this suggests a relationship between the vector spaces spanned by the columns of the design matrices of each model, which is a central element to any linear model. Hence, this condition on data structure casts the following questions.

1) Is there some sort of equivalence between the two-way Anova model and the Ancova model under this setting? 2) Is it possible to stablish a formal order relation between the precision of the two models? 3) If such relation exists, how the number of covariates impacts it? 4) What is the impact of adding covariates with such structure to the two-way Anova model?

In addition to these problems, an experiment performed to assess the effects of salinity on plant development leads to the same structure for the covariate named field capacity, which is related to the water content of the soil after water excess has been drained by gravity. In experimental conditions, this trait can be controlled via irrigation sheets, leading to a blocking factor defined by a quantitative variable, hence, for this covariate the experiment satisfies the same kind of restrictions discussed above, which were termed complete homogeneity (CH) conditions.

### *2.2 General statistical setting and notation*

In this section, the statistical models and the notation that will be used throughout the paper are introduced. As pointed out above, this study focuses on linear fixed effects models. The two-way Anova model without interaction (model 1) is as follows:

$$\boldsymbol{Y} = X\boldsymbol{\beta} + \boldsymbol{\varepsilon} = [T^* \vdots X_2]\begin{bmatrix}\boldsymbol{\theta}_1\\ \cdots\\ \boldsymbol{\beta}_2\end{bmatrix} + \boldsymbol{\varepsilon}$$

$$\boldsymbol{\varepsilon} \backsim N_n(\boldsymbol{0}, \sigma^2 I)$$

$$\Rightarrow \boldsymbol{Y} \sim N_n(X\boldsymbol{\beta}, \sigma^2 I)$$

where $T^* = [\boldsymbol{1}_n \vdots X_T], \dim(X_T) = n \times T, \dim(X_2) = n \times B,$ $T$ is the number of treatments, $B$ is the number of blocks, $n = \sum_{j=1}^{T}\sum_{l=1}^{B} m_{jl}$ is the sample size, $m_{jl}$ is the number of observations for the subclass formed by block $l$ and treatment $j$, $\boldsymbol{Y}$ is the random vector of responses, $\boldsymbol{\theta}_1$ is the vector containing the intercept and fixed treatment effects, $T^*$ is a matrix comprised of a vector of ones of dimension $n \times 1$ (denoted as $\boldsymbol{1}_n$) and the design matrix of treatment effects ($X_T$), $\boldsymbol{\beta}_2$ is the vector of fixed blocks effects, $X_2$ its design matrix, and $\boldsymbol{\varepsilon}$ the random vector of errors. On the other hand, the Ancova (model 2) is as follows:

$$\boldsymbol{Y} = W\boldsymbol{\theta} + \boldsymbol{\varepsilon} = [T^* \vdots W_2]\begin{bmatrix}\boldsymbol{\theta}_1\\ \cdots\\ \boldsymbol{\theta}_2\end{bmatrix} + \boldsymbol{\varepsilon}$$

$$\boldsymbol{\varepsilon} \sim N_n(\boldsymbol{0}, \sigma^2 I)$$

$$\Rightarrow \boldsymbol{Y} \sim N_n(W\boldsymbol{\theta}, \sigma^2 I)$$

where $\dim(W_2) = n \times p$, $p$ is the number of covariates, $\boldsymbol{\theta}_2$ is the vector of fixed regression coefficients of covariates, and $W_2$ is the corresponding design matrix. The remaining terms are as defined above.

Finally, consider the extension of model 1 to the two-way Anova model with interaction (model 3):

$$\boldsymbol{Y} = X^*\boldsymbol{\beta}^* + \boldsymbol{\varepsilon} = [T^* \vdots X_2 \vdots X_I]\begin{bmatrix}\boldsymbol{\theta}_1\\ \cdots\\ \boldsymbol{\beta}_2\\ \cdots\\ \boldsymbol{\beta}_I\end{bmatrix} + \boldsymbol{\varepsilon}$$

$$\boldsymbol{\varepsilon} \sim N_n(\boldsymbol{0}, \sigma^2 I)$$

$$\Rightarrow \boldsymbol{Y} \sim N_n(X^*\boldsymbol{\beta}^*, \sigma^2 I)$$

where, $\dim(X_I) = n \times BT$, and $X_I$ is the design matrix of block by treatment interaction effects.

**Some comments on notation**: For a matrix $A$, let $\langle A\rangle := Col(A)$ be its column space, that is, the vector space spanned by its columns. If $U$ and $V$ are vector spaces, then "$U \preccurlyeq V$" means that $U$ is a subspace of $V$; finaly $r(\cdot)$ is the rank of a matrix. Also, this notation applies for both: the randomized complete blocks design ($m_{lj} = 1\ \forall\ l\ \forall\ j$) and the generalized randomized complete blocks design ($m_{lj} \geq 2\ \forall\ l\ \forall\ j$).

### *2.3 Analytical model comparison*

Motivated by the problems described in section 2.1.2, we developed an analytical comparison of the models presented in section 2.2 in terms of precision, for the particular setting induced by this pattern on the covariates. The approach is based on the theory of orthogonal projections and vector (sub)spaces. In the first place, recall that under models 1, 2, and 3, precision can be assessed through the error variance $\sigma^2$ (Christensen, 2011). Let $SSM_X := \boldsymbol{Y}'P_X\boldsymbol{Y}$, $SSM_W := \boldsymbol{Y}'P_W\boldsymbol{Y}$, and $SSM_{X^*} := \boldsymbol{Y}'P_{X^*}\boldsymbol{Y}$ be the model sums of squares for models 1, 2 and 3, respectively, where $P_X$, $P_W$, and $P_{X^*}$ are the corresponding orthogonal projection matrices. Similarly, let $SSE_X := \boldsymbol{Y}'(I_n - P_X)\boldsymbol{Y}$, $SSE_W := \boldsymbol{Y}'(I_n - P_W)\boldsymbol{Y}$, and $SSE_{X^*} := \boldsymbol{Y}'(I_n - P_{X^*})\boldsymbol{Y}$ be the error sums of squares. Now, consider the following estimators of error variance for each model:

$$\hat{\sigma}_X^2 = \frac{SSE_X}{n - r(X)}, \hat{\sigma}_W^2 = \frac{SSE_W}{n - r(W)}, \hat{\sigma}_{X^*}^2 = \frac{SSE_{X^*}}{n - r(X^*)}$$

these are unbiased estimators corresponding to the mean square of the error (MSE) and are restricted maximum likelihood estimators as well. On the other hand, we have the maximum likelihood estimators (MLE):

$$\hat{\sigma}_{ML_X}^2 = \frac{SSE_X}{n}, \hat{\sigma}_{ML_W}^2 = \frac{SSE_W}{n}, \hat{\sigma}_{ML_{X^*}}^2 = \frac{SSE_{X^*}}{n}$$

When the problem is treated from a vector space point of view and the restrictions imposed by the settings described in section 2.1.2 are taken into account, some analytical solutions can be envisaged. From section 2.1.2, we have two conditions on the experimental design that lead to containment relations of the vector spaces $\langle X\rangle$, $\langle W\rangle$, and $\langle X^*\rangle$, that are useful to derive the theoretical results presented in this section.

**Condition CH1 (Complete Homogeneity 1)**: The $p$ covariates exhibit a pattern such that the value of each one of them is the same for all experimental/observational units pertaining to the same block.

**Condition CH2 (Complete Homogeneity 2)**: The $p$ covariates exhibit a pattern such that the value of each one of them is the same for all experimental/observational units in the same block-treatment subclass.

It is worth mentioning that albeit these conditions were motivated from problems in experimental design, they could be met in observational studies as well, moreover, models 1 to 3 can be used to analyse observational data. In the observational case, there is no need to talk about block and treatment, but factor 1 and factor 2. In addition, notice that a special case of these conditions is when the covariates are constant for all experimental/observational units, which of course, is of no interest and virtually impossible to observe in real-life studies. Consequently, whenever mentioning CH1 or CH2, this scenario is implicitly excluded.

Hereinafter, models such that the vector space spanned by the columns of their design matrix (estimation space) is the same, will be referred to as “equivalent”. Notice that, in this case, the model and error sums of squares, and the model and error degrees of freedom are the same. In particular, it implies that the two models yield the same precision. The following results summarize our formal findings.

**Theorem 1**. Under condition CH1 there exists a matrix $\Lambda$ with dimension $B \times p$ such that $W_2 = X_2\Lambda$, and $r(W_2) = r(\Lambda)$. In addition:

$i)$ $SSE_X \leq SSE_W$

$ii)$ $r(W) = \begin{cases} T + r(W_2) - 1, \text{if and only if, } \mathbf{1}_B \in \langle\Lambda\rangle. \\ T + r(W_2), \text{ if and only if, } \mathbf{1}_B \notin \langle\Lambda\rangle. \end{cases}$

$iii)$ $r(W) \leq T + B - 1$ with equality if: $r(\Lambda) = B$ or $r(\Lambda) = B - 1$ and $\mathbf{1}_B \notin \langle\Lambda\rangle$, in this case models 1 and 2 are equivalent.

$iv)$ If $r(\Lambda) = p$, which requires $p \leq B$, then linear functions of location parameters that are estimable under the one-way Anova are also estimable under model 2.

**Proof**. See Supplementary File 1.

**Theorem 2**. Let

$$\mathcal{H} := \left\{ H: H = X\Lambda = [T^* \vdots X_2] \begin{bmatrix} \Lambda_1 \\ \cdots \\ \Lambda_2 \end{bmatrix}, \Lambda_1 \neq \mathbf{0}, \Lambda_2 \neq \mathbf{0} \right\}$$

$i)$ Condition CH2 is necessary for $W_2 \subset \mathcal{H}$, and, in this case $SSE_X \leq SSE_W$.

$ii)$ $r(W_2) \leq B - 1 + \rho_I$ where $\rho_I := \dim(\langle T^* \rangle \cap \langle W_2 \rangle)$.

On the other hand, in the more general setting $W_2 \subset \langle X \rangle$ without the restriction $W_2 \subset \mathcal{H}$ it follows that:

$iii)$ The inequality $SSE_X \leq SSE_W$ still holds.

$iv)$ $W_2 \subset \langle T^* \rangle$ if and only if $\rho_I = r(W_2)$, in this case $r(W) = r(T^*)$ and model 2 is equivalent to the one-way Anova model.

$v)$ If $W_2 = X\Lambda$ with $\Lambda$ of full column rank and such that $\langle\Lambda\rangle$ and $\mathcal{N}(X)$ (the null space of $X$) are essentially disjoint, then $W_2$ is of full column rank and any function that is estimable in the one-way Anova model is also estimable in model 2.

$vi)$

$$\left\langle \begin{pmatrix} 1 \\ -\mathbf{1}_T \\ \mathbf{0}_{B\times 1} \end{pmatrix} \right\rangle \not\preccurlyeq \langle\Lambda\rangle \ or \ \left\langle \begin{pmatrix} -1 \\ \mathbf{0}_{T\times 1} \\ \mathbf{1}_B \end{pmatrix} \right\rangle \not\preccurlyeq \langle\Lambda\rangle$$

is a sufficient condition for $\langle\Lambda\rangle$ and $\mathcal{N}(X)$ to be essentially disjoint.

**Proof**. See Supplementary File 1.

**Theorem 3**. $i)$ Condition CH2 is necessary and sufficient for $W_2 \subseteq \langle X_I \rangle$.

$ii)$ Under condition CH2 $SSE_{X^*} \leq SSE_W$.

$iii)$ Let $\Lambda_{TB\times p}$ be the matrix such that $W_2 = X_I\Lambda$, and $Z_{TB\times T}$ be the matrix such that $X_T = X_I Z$. The following conditions are sufficient for models 2 and 3 to be equivalent:

a. The matrix $A := [Z \vdots -\Lambda]$ has full column rank and $r(\Lambda) = p = T(B-1)$.
b. $\Lambda$ reaches its maximum rank, that is, $r(\Lambda) = TB$.

$iv)$ If $\Lambda$ is of full column rank, which requires $p \leq TB$, then any function of treatment effects that is estimable under the one-way Anova is also estimable under model 3.

**Proof**: See Supplementary File 1.

These theorems set formal order relations between the ranks of the design matrices and the error sums of squares, as well as the impact of the CH structures on parameter estimability (as defined in the theory of linear models). Now that a theoretical foundation has been set, we can study some of its consequences on precision. The following corollary establishes how the number of covariates and the number of blocks impact order relations for model precision for a small sample size and a given number of treatments; in addition, it also shows the large sample behavior of model precision.

Notice that, when comparing model 1 to model 2 under condition CH1, if $r(W) < T + B - 1$ (i.e., models are not equivalent), then model 2 is a reduced version of model 1 because $\langle W \rangle \preccurlyeq \langle X \rangle$, thus, following Christensen (2011 section 3.2) it follows that if model 2 is correct, then model 1 is correct, conversely, if model 1 is correct, it does not necessarily imply that model 2 holds. Likewise, when comparing models 2 and 3, model 3 is the full model and model 2 is a reduced version, so the same implications follow. This discussion is relevant for part 2 of the following corollary.

**Corollary 1**.

**1.** *Small sample precision*. Under condition CH1 and $p \leq B - 1$, then

$$SSE_X(n - T - p) \leq SSE_W(n - T - B + 1)$$

is a necessary and sufficient condition for: $\hat{\sigma}_X^2 \leq \hat{\sigma}_W^2$. This also holds for the scenario discussed in the second part of Theorem 2 if $\Lambda_1$ is null.

**2.** *Large sample behavior of precision*. Consider $n \to \infty$.

2.1. Under condition CH1 or $W_2 \subset \langle X \rangle$; suppose that models 1 and 2 are not equivalent (see part $iii$ of Theorem 1). If model 2 is valid, which implies that $\hat{\sigma}_W^2$ is consistent, then, asymptotically, the models yield the same precision.

2.2. Under condition CH1 or $W_2 \subset \langle X \rangle$, if model 1 holds, but model 2 does not, for $n$ large enough, $\hat{\sigma}_X^2 < \hat{\sigma}_W^2$, disregarding if $\hat{\sigma}_W^2$ is a convergent sequence or not.

2.3. Parts 2.1 and 2.2 hold when comparing models 2 and 3, with model 3 playing the role of model 1 in 2.1 and 2.2, and replacing CH1 by CH2.

**3.** If condition CH1 holds, or $W_2 \subset \langle X \rangle$, then $\hat{\sigma}_{ML_X}^2 \leq \hat{\sigma}_{ML_W}^2$, also, if condition CH2 holds, then $\hat{\sigma}_{ML_{X^*}}^2 \leq \hat{\sigma}_{ML_W}^2$.

**Proof**. See Supplementary File 1.

Theorems 1 to 3 state conditions for linear combinations of treatment means that are estimable under the one-way Anova, to be estimable under models 2 and 3. This is of paramount importance in any application as the researchers are familiar with many kinds of such functions like pairwise mean differences and treatment means themselves. The following corollary is very relevant since it discusses precision of estimates for such functions. It is worth noticing that linear functions that are estimable in the one-way Anova are always estimable in model 1.

**Corollary 2**. Assume that the conditions stated in Theorems 1, 2 or 3 to guarantee that any estimable function in the one-way Anova is also estimable in models 2 or 3. Then, these functions are estimated with at least the same precision under model 1 than under model 2, and under model 3 than under model 2, provided that parts 1 or 2 of corollary 1 are fulfilled.

**Proof**. See Supplementary File 1.

A discussion regarding the discrepancy between the true mean of the response variable and its expectations under models 1 and 2 for the case of soil heterogeneity can be found in Supplementary File 2.

***2.4 Empirical model comparison***

This section describes a simulation study and the analysis of records from two field experiments to perform an empirical comparison of the performance of models 1 and 2 and to illustrate the insights obtained from section 2.3.

*2.4.1 Simulation study*

The main goal of this simulation was studying the performance of models 1 and 2 under complex soil heterogeneity patters induced by several soil properties when blocks are defined under conventional approaches, that is, in the absence of soil composition records using the standard partition into equally sized plots. This was done with the aim of finding out how much the performance of model 1 is lowered under such "adverse" scenario.

Thus, this simulation seeks to bring some insights regarding the research question posed in section 2.1.1

Using the premise that many variables of agronomical interest such as biomass production depend on soil properties (Thornley and France, 2007), this simulation modeled the response variable as a function of a collection of continuous variables representing soil characteristics, the treatment means and a random error. Data were simulated in two steps, the first one involved the simulation of soil characteristics and the second one the simulation of records as functions of three components: treatment means, a function of soil characteristics, and Gaussian independent errors.

Blocks were defined following the conventional approach which consists of splitting the experimental field (simulated as a rectangle) in equal (or approximately equal) plots. Three scenarios were defined depending on the number of blocks ($B = 3,4,6$); the simulation was repeated 500 times. Model performance was evaluated through: the squared root of mean squared error and mean absolute error of estimable functions (RMSEE and MAEE, respectively) corresponding to pairwise treatment differences, the squared root of mean squared error and mean absolute error computed from observed responses and fitted values of the response variable (RMSEy and MAEy, respectively), Akaike and Bayesian information criteria (AIC and BIC, respectively) and the accuracy of testing the hypothesis of equality of treatment means (HTAcc).

Because collecting and analyzing soil samples from all experimental/observational units may have a prohibitive economic cost, we considered missing covariates (soil properties). The typical case will be the one in which points to be sampled are defined at random, so $p$ soil variables are observed at $c$ out of $n$ points and, consequently, a missing at random scenario may be assumed. In order to assess the impact of missing data on the performance of model 2, for each simulated data set, a missing at random pattern was used to remove 10, 20, 30, 40, 50 or 60 % of the observations on each covariate. Our approach to deal with missing values was based on imputing missing records via kriging using an exponential variogram model as implemented in the Gstat library (Pedesma, 2004) of R (R Core Team, 2026). Once missing data were imputed, these were pooled with observed records to perform the analysis. Further details on data simulation may be found in Supplementary file 3.

*2.4.2 Real data analyses*

*Piglet performance experiment*. It was arranged in a generalized complete blocks design. The treatments design consisted of five diets fed to growing piglets from the same genetic line: a negative control (NC), a positive control, NC plus a full dose of probiotic 1, NC plus half dose of probiotic 1 and full dose of probiotic 2, and NC with full dose of probiotic 1 and probiotic 2. The blocking criteria was the production batch with three levels. Other details were already described in section 2.1.2. This dataset meets condition CH1.

*Radish experiment.* This study aimed at determining the effect of soil substratum salinity on radish (*Raphanus sativus*) development. The experiment was performed at the vegetal physiology greenhouse of the Agricultural Sciences Faculty of Universidad Nacional de Colombia, Bogotá, D.C., Colombia. The experiment was arranged in a generalized randomized complete blocks design with salinity as treatment and field capacity as block. Treatment design was unifactorial with four levels of salinity expressed as gr. of NaCl per lt. of water (0, 1.5, 3.0, and 4.5), whereas blocks corresponded to three irrigation sheets expressed as field capacity (60%, 80% and 100% field capacity). Each plant was the experimental unit, and there were five plants per block-treatment combination. The response variable was plant height (cm) and tuber weight (gr). This dataset meets condition CH1.

Records from both experiments were analyzed using standard procedures for Anova and Ancova models. All analyses were performed using an in-house R script (R Core Team, 2026) which is provided in Supplementary File 4.

## 3. Data analysis results

### *3.1 Simulation study*

Table 1 contains the summary of the simulation study for precision (as measured by the observed MSE), AIC, BIC, and the accuracy of testing the hypothesis of additive treatment effects. Also, Supplementary Figure 2 shows the spatial distribution of the four simulated soil variables for replicate 1.

It was found that model 2 outperformed model 1 in terms of precision as measured by MSE. Moreover, MSE increased as the percentage of missing data grew from 0 to 60%, but it was still smaller than the one obtained from model 1. Regarding the number of blocks, this parameter did not have a substantial impact on precision (Table 1). Finally, notice that MSE featured high variability across replicates as indicated by its coefficient of variation in all scenarios (Table 1).

**Table 1.** Across-replicates summary of the simulation study in different setups (the two models under different scenarios) for precision (as measured by MSE), information criteria and hypothesis testing accuracy.

| Setup[1] | Missing data % | MSE mean | MSE CV | AIC | BIC | HTAcc |
|---|---|---|---|---|---|---|
| Model 1 3B | -- | 46.61 | 0.80 | 161.75 | 169.99 | 0.81 |
| Model 1 4B | -- | 47.00 | 0.81 | 214.81 | 226.54 | 0.83 |
| Model 1 6B | -- | 46.42 | 0.76 | 319.50 | 338.21 | 0.89 |
| Model 2 NMC | 0 | 27.26 | 1.26 | 200.38 | 213.93 | 0.83 |
| Model 2 MC | 10 | 29.09 | 1.20 | 205.30 | 218.85 | 0.83 |
| Model 2 MC | 20 | 31.07 | 1.16 | 209.88 | 223.43 | 0.83 |
| Model 2 MC | 30 | 32.90 | 1.04 | 214.38 | 227.93 | 0.84 |
| Model 2 MC | 40 | 35.31 | 1.01 | 218.35 | 231.90 | 0.84 |
| Model 2 MC | 50 | 38.21 | 0.94 | 222.53 | 236.08 | 0.84 |
| Model 2 MC | 60 | 40.58 | 0.90 | 225.24 | 238.78 | 0.84 |

[1]For model 1, the number preceding the letter B indicates the number of blocks; NMC: no missing covariates; MC: Missing covariates.
MSE mean and MSE CV: mean and coefficient of variation (CV) of the estimated mean square of error, HTAcc: hypothesis test accuracy for equality of treatment means.

The information criteria (AIC and BIC) suggested a better performance of model 1 with three blocks followed by model 2 when there were no missing values, the highest (worst) values were attained under model 1 with six blocks. These criteria showed a poorer fit as the percentage of missing values increased (Table 1). On the other hand, results for the accuracy of testing the hypothesis of equal treatment means did not show marked differences; model 1 with six blocks had the best performance, whereas model 1 with three blocks had the worst; besides, in this case, increasing the percentage of missing data did not have an impact on performance, in fact, notice that the HTAcc increased slightly when going from 0% to 60% missing data.

Concerning the accuracy of the fitted values of the response variable and the estimates of pairwise treatment means differences as measured by MAEy and MAEE, respectively, model 2 showed a superior performance in MAEy, while MAEE was very similar across scenarios with a slightly better (smaller) result for model 2 with no missing values or a small percentage of missing values; moreover, both, MAEy and MAEE increased monotonically with the percentage of missing values (Supplementary table 1). As to the mean squared error, which combines accuracy and precision (Lehmann and Casella, 1998; Casella and Berger, 2002), it was found that model 2 outperformed model 1 under all missing data scenarios. Finally, in regard to RMSEE, the differences were not marked, but favored model 1 with six blocks.

### *3.1 Real dada Analysis*

Tables 2 and 3 show treatment and error sum of squares, mean squares, degrees of freedom and the F statistic for the hypothesis of additive treatment effects for the piglets and radish experiments, respectively, under the two-way Anova (without interaction) and Ancova models with different sets of covariates. As already noticed, both experiments meet condition CH1, so the observed results are consequences of Theorem 1, part 1 of corollary 1 and corollary 2, thus, we fitted one, two and three covariates.

**Table 2.** Relevant results from the analysis of variance for average body weight at 70 days (BW70) from the piglets' experiment using model 1 (two-way Anova without interaction) and model 2 (Ancova) with different covariates. Results from equivalent models are shown in bold

| Model[1] | Source | DF | SS | MS | F Stat. |
|---|---|---|---|---|---|
| **Two-way Anova** | **Diet** | **4** | **74.96** | **18.74** | **2.10** |
| | **Error** | **53** | **473.61** | **8.94** | |
| Ancova: Tmin | Diet | 4 | 74.96 | 18.74 | 2.13 |
| | Error | 54 | 474.66 | 8.79 | |
| **Ancova: TMin + MaxDeltaT** | **Diet** | **4** | **74.96** | **18.74** | **2.10** |
| | **Error** | **53** | **473.61** | **8.94** | |
| **Ancova: Tmax + TotPrec** | **Diet** | **4** | **74.96** | **18.74** | **2.10** |
| | **Error** | **53** | **473.61** | **8.94** | |
| **Ancova: TotPrec + MinPR + MinHR** | **Diet** | **4** | **74.96** | **18.74** | **2.10** |
| | **Error** | **53** | **473.61** | **8.94** | |

[1]For the Ancova models, abbreviations after the colon correspond to covariates included in the model. Tmin: minimum temperature, MaxDeltaT: maximum temperatura change, Tmax: máximum temperature, TotPrec: total precipitation, MinPR: minimum precipitation, MinHR: minimum relative humidity.

For all the studied variables, the treatment-block interaction was not significant; therefore, model 1 (two-way Anova without interaction) was fitted. Notice that, in Table 1, the condition stated in part 1 of corollary 1, namely, $SSE_X(n-T-p) \leq SSE_W(n-T-B+1)$, is not satisfied, hence, the Ancova model with temperature as covariate yields a slightly higher precision than the two-way Anova, thereby showing an example of necessity. On the other hand, in the radish experiment, tuber weight shows an example of sufficiency, because the condition was met and, consequently, the two-way Anova yielded higher precision, while plant height is another example of necessity. Moreover tables 2, 3 and supplementary table 2, show an example of part 3 of theorem 1, when $r(\Lambda) = B - 1$ and $\mathbf{1}_B \notin \langle \Lambda \rangle$. In Table 2, the Ancova model with two covariates is equivalent to model 1. Supplementary file 5 contains the R code two build matrix $\Lambda$ corresponding to the third model in Table 2, explicitly, in this case $r(\Lambda) = 2$ and $\mathbf{1}_3 \notin \langle\Lambda\rangle$ which explains the observed model equivalence. Besides, Tables 2 and 3 show a relevant consequence of our theoretical results, in these cases there is no gain in fitting more than 2 covariates, as it does not add any information or gain in precision, this is because $\langle W_2\rangle$ is a subspace of $\langle X_2\rangle$, and $r(X_2) = 3$ (the number of blocks), thus $r(W_2) \leq 3$, hence, when fitting 3 or more covariates $\langle X_2\rangle = \langle W_2\rangle$ and consequently, $\mathbf{1}_3 \in \langle\Lambda\rangle$ (see poof of Theorem 1) which implies that only the effects of two covariates are identifiable; for instance, when using the R program, after performing an alphanumeric ordering, the regression coefficient of the first covariate is set to zero. Furthermore, no matter which two covariates are fit, the equivalence holds. Thus, the practical implication for the piglet data is that albeit several environmental variables are available, any combination of three, yields a model equivalent to model 1, as well as any combination of two as far as $\mathbf{1}_3 \notin \langle\Lambda\rangle$. The main observation is that our theoretical results permit this information to be available in advance.

**Table 3**. Relevant results from the analysis of variance of data from the radish experiment using model 1 (two-way Anova without interaction) and model 2 (Ancova) with different covariates. Results from equivalent models are shown in bold.

| Response | Model[1] | Source | DF | SS | MS | F Stat. |
|---|---|---|---|---|---|---|
| Height | **Two-way Anova** | **Trt** | **3** | **1.0990** | **0.3663** | **0.2749** |
| | | **Error** | **54** | **71.9610** | **1.3326** | |
| | Ancova: Temp | Trt | 3 | 1.0990 | 0.3663 | 0.2788 |
| | | Error | 55 | 72.2810 | 1.3142 | |
| | **Ancova: Temp + Hum** | **Trt** | **3** | **1.0990** | **0.3663** | **0.2749** |
| | | **Error** | **54** | **71.9610** | **1.3326** | |
| Weight | **Two-way Anova** | **Trt** | **3** | **458.9200** | **152.9733** | **56.1332** |
| | | **Error** | **54** | **147.1600** | **2.7252** | |
| | Ancova: Temp | Trt | 3 | 458.9200 | 152.9733 | 53.5485 |
| | | Error | 55 | 157.1200 | 2.8567 | |
| | **Ancova: Temp + Hum** | **Trt** | **3** | **458.9200** | **152.9733** | **56.1332** |
| | | **Error** | **54** | **147.1600** | **2.7252** | |

[1]For the Ancova models, abbreviations after the colon correspond to covariates included in the model. Tmin: minimum temperature, MaxDeltaT: maximum temperatura change, Tmax: máximum temperature, TotPrec: total precipitation, MinPR: minimum precipitation, MinHR: minimum relative humidity.

## 4. Discussion

### *4.1 Blocking vs the use of covariates*

In agronomical studies, it is said that blocking loses efficiency if heterogeneity between plots does not follow a definite pattern (Yang and Juskiw, 2011) which is the case when certain soil variables follow complex patterns across the experimental field. Indeed, some results from our simulations provide empirical evidence supporting these ideas because it was found that this setting had a negative impact on precision for the two-way Anova model. However, the impact was null or small in terms of the accuracy of: 1) the fitted response, 2) the estimated pairwise differences between treatments means and 3) testing the hypothesis of additive treatment effects. The reasons behind these results are not clear and are matter for future work. The main insight from this simulation is that, under complex soil composition patterns, the traditional blocking approach may be inappropriate in terms of precision, but does not have a considerable impact on accuracy.

Moreover, simulations suggest that a relatively high proportion of missing values can be admitted as it did not show a substantial impact on accuracy and precision. This ability to deal with a high proportion of missing data makes Ancova analyses based on records from precision agriculture a promising alternative because it allows its implementation at a lower cost. Another alternative to reduce the implementation cost is using variables that somehow summarize relevant soil properties and are cheaper to measure. In that regard, Johnson, Eskridge and Corwin (2005) discussed the usefulness of using soil apparent electrical conductivity to define blocks in large- and small-scale experiments.

In our simulation study, the concomitant variables were simulated to emulate spatial correlation and to exhibit complex variation patterns across the experimental field (Supplementary Figure 2); therefore, there is no within-block homogeneity. Albeit this fact, it is striking that in contrast to the results from Klockars, Potter and Beretvas (1999), in our simulation study, the accuracy of testing for additive treatment effects of a model considering block effects was equal or slightly superior to that of Ancova.

On the other hand, some authors recommend the use of both, block, and covariates effects because the concomitant variables may remove spatial variability that is not controlled by blocks (Schewenke, 1997; Yang and Juskiw, 2011). This is an interesting approach that combines model 1 and model 2 or model 2 and model 3. However, under the CH conditions studied here, this kind of model makes no sense because the part of the design matrix corresponding to covariates is spanned by the columns of $X$, $X_I$, or a proper subset of them, consequently, under any of these settings there is no gain in a model considering block and covariates effects because the rank of the design matrix would remain the same.

### *4.2 Further considerations*

Theorems 1, 2 and 3, along with corollaries 1 and 2, set a baseline for the use of vector spaces and orthogonal projections in a formal comparison of the two-way Anova and Ancova models, two of the most frequently used statistical models in applied research. It is worth emphasizing that no similar formulations (i.e., based on orthogonal projections and the theory of vector spaces) were found in the literature; therefore, the approach itself may be useful as a tool to prove similar results in other scenarios or to extend our findings. Also, because these are algebraic results, they do not rely on probabilistic assumptions.

It is worth noticing that under the CH conditions and for small sample sizes, it is still possible for the Ancova model (model 1) to be more efficient than the two-way Anova without interaction (model 1), as far as the condition in part 1 of corollary 1 is not fulfilled, as discussed in section 3.1 (Table 2). Similarly, when considering the two-way Anova with interaction (model 3), the Ancova may be more efficient. This phenomenon is due to the trade-off between a larger SSE and a larger number of error degrees of freedom exhibited by the reduced models.

In addition to CH1 and CH2, other conditions stated in Theorems 1, 2, 3, and their corollaries must be satisfied to obtain more precise estimates of estimable functions. A key condition to guarantee that standard parametric functions of treatment effects that are estimable under the one-way Anova, are also estimable under models 2 and 3 is that $\Lambda$ is of full column rank (recall that $\Lambda$ is not the same matrix in theorems 1, 2 and 3). Checking this condition is a standard procedure in linear algebra (see Supplementary file 5); however, it is worth mentioning a particular case of pragmatic interest. Suppose that there is a single covariate satisfying CH1; therefore, $\Lambda$ is of full column rank and, consequently, if $SSE_X(n-T-p) \leq SSE_W(n-T-B+1)$ (corollary 1 part 1) or for $n$ large enough (corollary 1 part 2), model 1 yields estimates of estimable functions of treatment means at least as precise as those from model 2. Further, note that $\mathbf{1}_B \notin \langle \Lambda \rangle$, thus, if there are two blocks, the models are equivalent, vide Theorem 1. Similar results follow for theorem 3. Moreover, in addition to $\Lambda$ being of full column rank, Theorem 2 requires one more condition to satisfy the aforementioned estimability guarantee, namely that $\langle \Lambda \rangle$ and $\mathcal{N}(X)$ are essentially disjoint. From Theorem 2, $\left\langle \begin{pmatrix} 1 \\ -\mathbf{1}_T \\ \mathbf{0}_{B\times 1} \end{pmatrix} \right\rangle \not\preccurlyeq \langle \Lambda \rangle \; or \; \left\langle \begin{pmatrix} -1 \\ \mathbf{0}_{T\times 1} \\ \mathbf{1}_B \end{pmatrix} \right\rangle \not\preccurlyeq \langle \Lambda \rangle$ is a sufficient condition. Clearly, when there is a single covariate which takes only positive values (as most environmental and soil variables), this sufficient condition is met. Supplementary file 5 provides an R code showing how to find $\Lambda$ and computing its rank, along with illustrations of our theoretical results for the piglets dataset.

## 5 Final Remarks

Data structures induced by the CH conditions coupled with settings stated in theorems 1 to 3, lead to the conclusion that observing more covariates than the number of blocks (one observation per block-treatment subclass) or the number of block-treatment subclasses (two or more observations per block-treatment subclass), does not increase the precision of the two-way Anova (i.e., local control through blocking); therefore, it gives a practical recommendation for the maximum number of covariates to be measured. Besides, for the problem of controlling for soil heterogeneity, so far, there are no general conditions guaranteeing a superior performance of a particular model. Our empirical results suggest that when spatial variability of soil features is complex, blocks defined with the traditional naïve approach undermines efficiency, but it seems that it does not affect hypothesis testing accuracy for additive treatment effects. Finally, with the widespread adoption of precision agriculture, records derived from these programs may be used as an alternative to perform local control of soil heterogeneity in agronomical research.

## References

Behera, S. K., Mathur, R. K., Shukla, A. K., Suresh, K., & Prakash, C. (2018). Spatial variability of soil properties and delineation of soil management zones of oil palm plantations grown in a hot and humid tropical region of southern India. *Catena*, *165*(February), 251–259. https://doi.org/10.1016/j.catena.2018.02.008

Casella, G., & Berger, R. L. (2002). Statistical Inference. (2nd ed). Duxbury.

Christensen, R. (2011). *Plane Answers to Complex Questions. The Theory of Linear Models* (4th ed.). Springer.

Cox, D. R. (1957). The Use of a Concomitant Variable in Selecting an Experimental Design. *Biometrika*, *44*(1/2), 150. https://doi.org/10.2307/2333247

Dempster, A. P., Laird, N. M., & Rubin, D. B. (1977). Maximum Likelihood from Incomplete Data Via the EM Algorithm. *Journal of the Royal Statistical Society: Series B (Methodological)*, *39*(1), 1–22. https://doi.org/10.1111/j.2517-6161.1977.tb01600.x

Federer, W. T. (1955). *Experimental design, theory and application*. Macmillan.

Feldt, L. S. (1958). A comparison of the precision of three experimental designs employing a concomitant variable. *Psychometrika*, *23*(4), 335–353. https://doi.org/10.1007/BF02289783

Florinsky, I. V. (2016). *Digital Terrain Analysis in Soil Science and Geology* (2nd ed.). Academic Press.

Goovaerts, P. (1998). Geostatistical tools for characterizing the spatial variability of microbiological and physico-chemical soil properties. *Biology and Fertility of Soils*, *27*(4), 315–334. https://doi.org/10.1007/s003740050439

Hong, N., White, J. G., Gumpertz, M. L., & Weisz, R. (2005). Spatial analysis of precision agriculture treatments in randomized complete blocks: Guidelines for covariance model selection. *Agronomy Journal*, *97*(4), 1082–1096. https://doi.org/10.2134/agronj2003.0130

Johnson, C. K., Eskridge, K. M., & Corwin, D. L. (2005). Apparent soil electrical conductivity: Applications for designing and evaluating field-scale experiments. *Computers and Electronics in Agriculture*, *46*(1-3 SPEC. ISS.), 181–202. https://doi.org/10.1016/j.compag.2004.12.001

Klockars, A. J., & Beretvas, S. N. (2001). Analysis of covariance and randomized block design with heterogeneous slopes. *Journal of Experimental Education*, *69*(4), 393–410. https://doi.org/10.1080/00220970109599494

Klockars, A. J., Potter, N. S., & Beretvas, S. N. (1999). Power to detect additive treatment effects with randomized block and analysis of covariance designs. *Journal of Experimental Education*, *67*(2), 180–191. https://doi.org/10.1080/00220979909598352

Kuehl, R. O. (2001). *Design of Experiments: Statistical Principles of Research Design and Analysis* (2nd ed.). Duxbury Press.

Lehmann, E. L., & Casella, G. (1998). Theory of Point Estimation  (2 ed). Springer Texts in Statistics. In *Design* (Vol. 41, Issue 3). https://doi.org/10.2307/1270597

Melo, O. O., López, L. A., & Melo, S. E. (2020). *Diseño de Experimentos: Métodos y Aplicaciones* (2nd ed.). Universidad Nacional de Colombia. https://repositorio.unal.edu.co/handle/unal/79912

Pebesma, E.J. (2004). Multivariable geostatistics in S: The gstat package. *Computers & geosciences*, *30(7)*, 683-691.

R Core Team. (2026). R: *A language and environment for statistical computing*. R Foundation  for Statistical Computing, Vienna, Austria. URL https://www.R-project.org/.

Schwenke, J. R. (1997). Comparing the Use of Block and Covariate Information in Analysis of Variance. *Conference on Applied Statistics in Agriculture*. https://doi.org/10.4148/2475-7772.1299

Stroup, W. W., Baenziger, P. S., & Multize, D. K. (1994). Removing spatial variation from wheat yield trials: A comparison of methods. *Crop Science*, *34*(1), 62–66. https://doi.org/10.2135/cropsci1994.0011183X003400010011x

Thornley, J. H. M.; France, J. (2007). *Mathematical models in agriculture: quantitative methods for the plant, animal and ecological sciences* (J. H. M. Thornley & J. France (eds.); (2nd ed.). CABI. https://doi.org/10.1079/9780851990101.0000

Vanden Heuvel, R. M. (1996). The promise of precision agriculture. *Journal of Soil and Water Conservation*, *51*(1), 38–40.

Wu, Y.-C., & McLean, J. E. (1994). To block or covary a concomitant variable: Which is more powerful? *Annual Meeting of the Mid-South Educational Research Association*, *November*, 28.

Yang, R. C., & Juskiw, P. (2011). Analysis of covariance in agronomy and crop research. *Canadian Journal of Plant Science*, *91*(4), 621–641. https://doi.org/10.4141/cjps2010-032

Zhang, N., Wang, M., & Wang, N. (2002). Precision agriculture - A worldwide overview. *Computers and Electronics in Agriculture*, *36*(2–3), 113–132. https://doi.org/10.1016/S0168-1699(02)00096-0

**Supplementary file 1.**
**Proofs of theorems and corollaries.**

**Theorem 1.** Under condition CH1 there exists a matrix $\Lambda_{B\times p}$ such that $W_2 = X_2\Lambda$, with $r(W_2) = r(\Lambda)$, and:
$i)$ $SSE_X \le SSE_W$
$ii)$ $r(W) = \begin{cases} T + r(W_2) - 1, \text{if, and only if, } \mathbf{1}_B \in \langle\Lambda\rangle. \\ T + r(W_2), \text{if, and only if, } \mathbf{1}_B \notin \langle\Lambda\rangle. \end{cases}$
$iii)$ $r(W) \le T + B - 1$ with equality if: $r(\Lambda) = B$ or $r(\Lambda) = B - 1$ and $\mathbf{1}_B \notin \langle\Lambda\rangle$, in this case models 1 and 2 are equivalent.
$iv)$ If $r(\Lambda) = p$, which requires $p \le B$, then linear functions of location parameters that are estimable under the one-way Anova model are also estimable under model 2.
**Proof.**
Due to the restriction imposed by condition CH1, it follows that each one of the columns of $W_2$ can be written as a linear combination of the columns of the part of the design matrix corresponding to block effects, that is, $X_2$. This ensures that the value of the covariate is the same for all experimental units pertaining to the same block; therefore, there exists a matrix $\Lambda_{B\times p}$ such that $W_2 = X_2\Lambda$. Moreover, since $X_2$ is of full column rank, it follows that:

$$r(W_2) = r(\Lambda).$$

$i)$ Let $SSM_X \coloneqq y'P_Xy$ and $SSM_W \coloneqq y'P_Wy$ be the model sums of squares of model 1 and model 2, respectively, where $P_X$ and $P_W$ are the corresponding orthogonal projection matrices onto $\langle X\rangle$ and $\langle W\rangle$. Notice that by properties of vector subspaces:

$$\langle X\rangle = \langle[T^* \vdots X_2]\rangle = \langle T^* \cup X_2\rangle = \langle T^*\rangle + \langle X_2\rangle \qquad (1)$$

where $\langle T^*\rangle + \langle X_2\rangle$ refers to the sum of subspaces, similarly

$$\langle W\rangle = \langle T^*\rangle + \langle W_2\rangle \qquad (2)$$

because $\langle W_2\rangle \preccurlyeq \langle X_2\rangle$, by (1) and (2) it follows that $\langle W\rangle \preccurlyeq \langle X\rangle$. From this fact and the properties $\langle P_X\rangle = \langle X\rangle$ and $\langle P_W\rangle = \langle W\rangle$, applying theorem B.47 in Christensen (2011) we have that $P_X - P_W$ is an orthogonal projector and, consequently, it is a non-negative definite matrix which induces:

$$\boldsymbol{y}'(P_X - P_W)\boldsymbol{y} \ge 0$$
$$\Leftrightarrow \boldsymbol{y}'P_X\boldsymbol{y} \ge \boldsymbol{y}'P_W\boldsymbol{y}$$
$$\Leftrightarrow \boldsymbol{y}'(I - P_X)\boldsymbol{y} \le \boldsymbol{y}'(I - P_W)\boldsymbol{y}$$

that is,

$$SSE_X \le SSE_W.$$

$ii)$ We use the following lemma.

**Lemma 1**. $X_2\boldsymbol{\rho} = \mathbf{1}_n$ if, and only if, $\boldsymbol{\rho} = \mathbf{1}_B$.
**Proof**. Let $\mathbf{1}_k, \mathbf{0}_k, k \in \mathbb{N}$, be k-dimensional column vectors whose entries are all equal to 1 and 0, respectively, also let $n_{1\cdot}, n_{2\cdot}, \dots, n_{B\cdot}$ be the number of records per block. Without loss of generality, data can be ordered by block, then $X_2$ has the form:

$$X_2 = \begin{bmatrix} \mathbf{1}_{n_{1\cdot}} & \mathbf{0}_{n_{1\cdot}} & \cdots & \mathbf{0}_{n_{1\cdot}} \\ \mathbf{0}_{n_{2\cdot}} & \mathbf{1}_{n_{2\cdot}} & \cdots & \mathbf{0}_{n_{2\cdot}} \\ \vdots & \vdots & \ddots & \vdots \\ \mathbf{0}_{n_{B\cdot}} & \mathbf{0}_{n_{B\cdot}} & \cdots & \mathbf{1}_{n_{B\cdot}} \end{bmatrix}$$

Thus, if $\boldsymbol{\rho} = \mathbf{1}_B$ the matrix product $X_2\boldsymbol{\rho}$ corresponds to the sum of the columns of $X_2$; therefore, it is immediate that $X_2\boldsymbol{\rho} = \mathbf{1}_n$. On the other hand, let $\boldsymbol{\rho} = (\rho_1\ \rho_2 \cdots \rho_B)'$ and consider the linear system (in $\boldsymbol{\rho}$) $X_2\boldsymbol{\rho} = \mathbf{1}_n$ which has explicit form:

$$\begin{bmatrix} \rho_1 \mathbf{1}_{n_{1\cdot}} \\ \rho_2 \mathbf{1}_{n_{2\cdot}} \\ \vdots \\ \rho_B \mathbf{1}_{n_{B\cdot}} \end{bmatrix} = \begin{bmatrix} \mathbf{1}_{n_{1\cdot}} \\ \mathbf{1}_{n_{2\cdot}} \\ \vdots \\ \mathbf{1}_{n_{B\cdot}} \end{bmatrix}$$

whose unique solution is $\rho_1 = \rho_2 = \cdots = \rho_B = 1$, that is, $\boldsymbol{\rho} = \mathbf{1}_B$.

**Q.E.D.**

Recall that the rank of a matrix, denoted as $r(\cdot)$, is the dimension of its column space, thus, because of the containment relationship $\langle W \rangle \preccurlyeq \langle X \rangle$, it follows that $r(W) \leq r(X)$. Now, for the two-way Anova model without interactions it is known that $r(X) = T + B - 1$ and following results from the theory of vector spaces the general expression for $r(W)$ is:

$$r(W) = \dim(\langle W \rangle) = \dim(\langle T^* \rangle) + \dim(\langle W_2 \rangle) - \dim(\langle T^* \rangle \cap \langle W_2 \rangle).$$

The following results will be used: $\exists\ \Lambda_{B\times p} \ni W_2 = X_2\Lambda$, $\langle T^* \rangle \cap \langle X_2 \rangle = \langle \mathbf{1}_n \rangle$, and the sum of the columns of $X_2$ is the vector $\mathbf{1}_n$, that is, $X_2\mathbf{1}_B = \mathbf{1}_n$.
If $\mathbf{1}_B \in \langle \Lambda \rangle$ then:

$$\begin{gathered} \exists \boldsymbol{\delta} \ni \Lambda\boldsymbol{\delta} = \mathbf{1}_B \\ \Longrightarrow X_2\Lambda\boldsymbol{\delta} = \mathbf{1}_n \\ \Longrightarrow W_2\boldsymbol{\delta} = \mathbf{1}_n \\ \Longrightarrow \mathbf{1}_n \in \langle W_2 \rangle \end{gathered}$$

and due to the fact that $\mathbf{1}_n \in \langle T^* \rangle$ then $\mathbf{1}_n \in \langle T^* \rangle \cap \langle W_2 \rangle$ which implies that

$$\langle \mathbf{1}_n \rangle \preccurlyeq \langle T^* \rangle \cap \langle W_2 \rangle$$

but $\langle W_2 \rangle \preccurlyeq \langle X_2 \rangle$ so $\langle T^* \rangle \cap \langle W_2 \rangle \preccurlyeq \langle T^* \rangle \cap \langle X_2 \rangle = \langle \mathbf{1}_n \rangle$ which shows that:

$$\begin{gathered} \langle T^* \rangle \cap \langle W_2 \rangle = \langle \mathbf{1}_n \rangle \\ \Longrightarrow \dim(\langle T^* \rangle \cap \langle W_2 \rangle) = 1 \end{gathered}$$

and consequently $r(W) = T + \dim(\langle W2 \rangle) - 1$.

Conversely, if $r(W) = T + \dim(\langle W2 \rangle) - 1$, from the general expression for $r(W)$ written above, the following equality holds:

$$T + \dim(\langle W_2 \rangle) - 1 = \dim(\langle T^* \rangle) + \dim(\langle W_2 \rangle) - \dim(\langle T^* \rangle \cap \langle W_2 \rangle)$$
$$\Leftrightarrow T - 1 = T - \dim(\langle T^* \rangle \cap \langle W_2 \rangle) \ (\because \dim(\langle T^* \rangle) = T)$$
$$\Leftrightarrow \dim(\langle T^* \rangle \cap \langle W_2 \rangle) = 1 \quad (3)$$

Again, $\langle W_2 \rangle \preccurlyeq \langle X_2 \rangle$ implies $\langle T^* \rangle \cap \langle W_2 \rangle \preccurlyeq \langle T^* \rangle \cap \langle X_2 \rangle = \langle \mathbf{1}_n \rangle$, a vector space of dimension 1, so by (3) it follows that $\langle T^* \rangle \cap \langle W_2 \rangle = \langle \mathbf{1}_n \rangle$, then, in particular, $\mathbf{1}_n \in \langle W_2 \rangle$ and, therefore:

$$\exists\ \boldsymbol{\delta} \ni W_2 \boldsymbol{\delta} = \mathbf{1}_n = X_2 \Lambda \boldsymbol{\delta}$$

but, by lemma 1, it implies that $\Lambda \boldsymbol{\delta} = \mathbf{1}_B$, which means $\mathbf{1}_B \in \langle \Lambda \rangle$. Consequently, we conclude that $r(W) = T + \dim(\langle W_2 \rangle) - 1\ iff\ \mathbf{1}_B \in \langle \Lambda \rangle$.

Now, let's prove the second part of $ii$. To this ends we use the following lemma.

**Lemma 2**. If $W_2 = X_2 \Lambda$, then $\mathbf{1}_B \in \langle \Lambda \rangle$ if, and only if, $\mathbf{1}_n \in \langle W_2 \rangle$.
**Proof.** If $\mathbf{1}_B \in \langle \Lambda \rangle \Rightarrow \exists\ \delta \ni\ \Lambda \delta = \mathbf{1}_B \Rightarrow W_2 \delta = X_2 \Lambda \delta = X_2 \mathbf{1}_B = \mathbf{1}_n \Rightarrow \mathbf{1}_n \in \langle W_2 \rangle$.
The implication in the other direction is proved by contradiction. Suppose that $\mathbf{1}_n \in \langle W_2 \rangle$, but $\mathbf{1}_B \notin \langle \Lambda \rangle$, then $\nexists\ \boldsymbol{\gamma} \in \mathbb{R}^p$ such that $\Lambda \boldsymbol{\gamma} = \mathbf{1}_B$. Thus, since $\mathbf{1}_n \in \langle W_2 \rangle$

$$\exists\ \boldsymbol{\delta} \ni\ W_2 \boldsymbol{\delta} = \mathbf{1}_n$$
$$\Rightarrow (X_2 \Lambda) \boldsymbol{\delta} = \mathbf{1}_n \ (\because\ W_2 = X_2 \Lambda)$$
$$\Rightarrow X_2 (\Lambda \boldsymbol{\delta}) = \mathbf{1}_n$$

but this implies that $X_2 \boldsymbol{b} = \mathbf{1}_n$ with $\boldsymbol{b} \coloneqq \Lambda \boldsymbol{\delta} \in \mathbb{R}^B, \boldsymbol{b} \neq \mathbf{1}_B$, which contradicts lemma 1; therefore, if $\mathbf{1}_n \in \langle W_2 \rangle \Rightarrow \mathbf{1}_B \in \langle \Lambda \rangle$; this completes the proof.

**Q.E.D.**

Resuming the proof of the second part of $ii$), if $\mathbf{1}_B \notin \langle \Lambda \rangle \Rightarrow \mathbf{1}_n \notin \langle W_2 \rangle$ (lemma 2). Now, once again we use the fact that $\langle T^* \rangle \cap \langle W_2 \rangle \preccurlyeq \langle T^* \rangle \cap \langle X_2 \rangle = \langle \mathbf{1}_n \rangle$, but $\mathbf{1}_n \notin \langle W_2 \rangle$, then

$$\mathbf{1}_n \notin \langle T^* \rangle \cap \langle W_2 \rangle \Rightarrow \langle T^* \rangle \cap \langle W_2 \rangle = \mathbf{0}_n.$$

Consequently, $\dim(\langle T^* \rangle \cap \langle W_2 \rangle) = 0$ and, using the general expression for $r(W)$:

$$r(W) = \dim(\langle T^* \rangle) + \dim(\langle W_2 \rangle) - \dim(\langle T^* \rangle \cap \langle W_2 \rangle)$$
$$= T + \dim(\langle W_2 \rangle).$$

Conversely, if $r(W) = T + \dim(\langle W_2 \rangle)$, using the general expression for $r(W)$ again:

$$T + \dim(\langle W_2 \rangle) - \dim(\langle T^* \rangle \cap \langle W_2 \rangle) = T + \dim(\langle W_2 \rangle)$$
$$\Rightarrow \dim(\langle T^* \rangle \cap \langle W_2 \rangle) = 0$$
$$\Rightarrow \langle T^* \rangle \cap \langle W_2 \rangle = \mathbf{0}_n$$
$$\Rightarrow \mathbf{1}_n \notin \langle W_2 \rangle \ (\because\ \mathbf{1}_n \in \langle T^* \rangle)$$
$$\Rightarrow \mathbf{1}_B \notin \langle \Lambda \rangle \ (\because lemma\ 2).$$

$iii$) Since $\langle W_2 \rangle \preccurlyeq \langle X_2 \rangle$, then:

$$r(W_2) = \dim(\langle W_2 \rangle) \leq \dim(\langle X_2 \rangle) = r(X_2) = B.$$

From part $ii$, $r(W)$ depends on the disjoint events $\mathbf{1}_B \in \langle \Lambda \rangle$, $\mathbf{1}_B \notin \langle \Lambda \rangle$. If $\mathbf{1}_B \in \langle \Lambda \rangle$ then

$$r(W) = T + r(W_2) - 1 \leq T + B - 1.$$

Now, notice that if $r(W_2) = B$ then $\dim(\langle W_2 \rangle) = \dim(\langle X_2 \rangle)$, but $W_2 \subseteq \langle X_2 \rangle$; therefore, $\langle W_2 \rangle = \langle X_2 \rangle$, but $\mathbf{1}_n \in \langle X_2 \rangle$, consequently, in this case $\mathbf{1}_n \in \langle W_2 \rangle$ and, by lemma 2, $\mathbf{1}_B \in \langle \Lambda \rangle$, then, by part $ii$: $r(W) = T + r(W_2) - 1 = T + B - 1$, this shows the sufficiency of $r(W_2) = B$ for the equality. Note that we just show that $r(W_2) = B \Rightarrow \mathbf{1}_B \in \langle \Lambda \rangle$, which implies that $\mathbf{1}_B \notin \langle \Lambda \rangle \Rightarrow r(W_2) \neq B$; however, $r(W_2) \leq B$, which means $\mathbf{1}_B \notin \langle \Lambda \rangle \Rightarrow r(W_2) < B$, but $r(W_2)$ is a natural number, hence, the last inequality is equivalent to $r(W_2) \leq B - 1$. Using this result, and part $ii$ it follows that if $\mathbf{1}_B \notin \langle \Lambda \rangle$ then

$$r(W) = T + r(W_2) \leq T + B - 1.$$

Consequently, disregarding if $\mathbf{1}_B$ is in the column space of $\Lambda$ or not, in general:

$$r(W) \leq T + B - 1.$$

If $\mathbf{1}_B \notin \langle \Lambda \rangle$ and $r(W_2) = B - 1$, using part $ii$ again: $r(W_2) = T + r(W_2) = T + B - 1$, this shows the sufficiency of $r(W_2) = B - 1, \mathbf{1}_B \notin \langle \Lambda \rangle$ for the equality.

Finally, recalling that $\langle W \rangle \preccurlyeq \langle X \rangle$, if $r(W) = T + B - 1$, then $r(W) = r(X)$, hence, model 1 and model 2 are equivalent as defined in the manuscript.

$iv$) Recall that $r(W_2) = r(\Lambda)$ and $r(W_2) \leq B$; moreover $\dim(\Lambda) = B \times p$, which implies that $r(\Lambda) \leq \min\{B, p\}$. Thus, if $p > B$, $W_2$ cannot be of full column rank, hence the necessity of the condition $p \leq B$. If $r(\Lambda) = p$, following the results shown in Christensen (2011) pg. 216-219, it follows that those parametric functions of location parameters that are estimable under the one-way anova model are also estimable under the ancova model (model 2).

**Q.E.D**

**Theorem 3**. $i$) Condition CH2 is necessary and sufficient for $W_2 \subseteq \langle X_I \rangle$.
$ii$) Under condition CH2 $SSE_{X^*} \leq SSE_W$.
$iii$) Let $\Lambda_{TB \times p}$ be the matrix such that $W_2 = X_I \Lambda$, and $Z_{TB \times T}$ be the matrix such that $X_T = X_I Z$. The following conditions are sufficient for models 2 and 3 to be equivalent:

a. The matrix $A := [Z \vdots -\Lambda]$ has full column rank and $r(\Lambda) = p = T(B - 1)$.
b. $\Lambda$ reaches its maximum rank, that is, $r(\Lambda) = TB$.

$iv$) If $\Lambda$ is of full column rank, which requires $p \leq TB$, then any function of treatment effects that is estimable under the one-way Anova is also estimable under model 2.

**Proof.**

***i***) *Sufficiency*: If condition CH2 holds, then, any covariate has the same value within each block-treatment subclass; therefore, when constructing the design matrix for covariates ($W_2$) it has the form of an element of the vector space spanned by the columns of

$$X^* = [T^* \vdots X_2 \vdots X_I],$$

the design matrix of the two-way Anova model with interaction, but $[T^* \vdots X_2] \subset \langle X_I \rangle$ which implies that $\langle X^* \rangle = \langle X_I \rangle \Rightarrow W_2 \subseteq \langle X_I \rangle$.
*Necessity*: If $W_2 \subseteq \langle X_I \rangle$, then $\exists\ \Lambda_{BT\times p} \ni\ W_2 = X_I\Lambda$. Let $n_{11}, n_{12}, \dots, n_{BT}$ be the number of records per block-treatment combination, and:

$$\Lambda = \begin{bmatrix} \lambda_{11} & \lambda_{12} & \cdots & \lambda_{1p} \\ \lambda_{21} & \lambda_{22} & \cdots & \lambda_{2p} \\ \vdots & \vdots & \ddots & \vdots \\ \lambda_{(BT)1} & \lambda_{(BT)2} & \cdots & \lambda_{(BT)p} \end{bmatrix}$$

Without loss of generality, data can be ordered by block and treatment, then $W_2$ has the form:

$$W_2 = \begin{bmatrix} \lambda_{11}\mathbf{1}_{n_{11}} & \lambda_{12}\mathbf{1}_{n_{11}} & \cdots & \lambda_{1p}\mathbf{1}_{n_{11}} \\ \lambda_{21}\mathbf{1}_{n_{12}} & \lambda_{22}\mathbf{1}_{n_{12}} & \cdots & \lambda_{2p}\mathbf{1}_{n_{12}} \\ \vdots & \vdots & \ddots & \vdots \\ \lambda_{(BT)1}\mathbf{1}_{n_{BT}} & \lambda_{(BT)2}\mathbf{1}_{n_{BT}} & \cdots & \lambda_{(BT)p}\mathbf{1}_{n_{BT}} \end{bmatrix}$$

which translates into condition CH2.

***ii***) Using the fact that $\langle X^* \rangle = \langle X_I \rangle$, the design matrices of models 2 and 3 satisfy the relationship: $\langle W_2 \rangle \preccurlyeq \langle X_I \rangle$. Thus, this proof is completely analogous to that of part $i$ of Theorem 1.

***iii***) a. Recall that

$$\begin{aligned} r(W) &= r(T^*) + r(W_2) - \dim(\langle T^* \rangle \cap \langle W_2 \rangle) \\ &= T + r(\Lambda) - \dim(\langle T^* \rangle \cap \langle W_2 \rangle) \end{aligned}$$

Now, notice that

$$\langle T^* \rangle \cap \langle W_2 \rangle \coloneqq \{\boldsymbol{u} \in \mathbb{R}^n : \exists\ \boldsymbol{a} \in \mathbb{R}^T, \boldsymbol{b} \in \mathbb{R}^p \ni \boldsymbol{u} = X_T\boldsymbol{a} = W_2\boldsymbol{b}\}$$

but $W_2 = X_I\Lambda$, and $X_T = X_I Z$, thus:

$$\begin{aligned} X_T\boldsymbol{a} = W_2\boldsymbol{b} &\Leftrightarrow X_I Z\boldsymbol{a} = X_I\Lambda\boldsymbol{b} \\ &\Leftrightarrow X_I(Z\boldsymbol{a} - \Lambda\boldsymbol{b}) = \mathbf{0}_n \\ &\Leftrightarrow Z\boldsymbol{a} - \Lambda\boldsymbol{b} = \mathbf{0}_{BT}\ (\because X_I\ is\ of\ full\ column\ rank) \\ &\Leftrightarrow [Z \vdots -\Lambda]\begin{bmatrix} \boldsymbol{a} \\ \cdots \\ \boldsymbol{b} \end{bmatrix} = \mathbf{0}_{BT} \end{aligned} \tag{13}$$

If $A_{BT\times(T+p)} \coloneqq [Z \vdots -\Lambda]$ has full column rank (which necessitates $Z$ and $\Lambda$ being of full column rank, recall that $\langle\Lambda\rangle$ and $\langle-\Lambda\rangle$ are isomorphic vector spaces), then the homogeneous linear system (13) has $\mathbf{0}_{T+p}$ as its unique solution, which implies that $\boldsymbol{a} = \mathbf{0}_T, \boldsymbol{b} = \mathbf{0}_p$ are the only vectors satisfying $\boldsymbol{u} = X_T\boldsymbol{a} = W_2\boldsymbol{b}$, hence $\langle T^*\rangle \cap \langle W_2\rangle = \{\mathbf{0}_n\}$; consequently, it follows that:

$$\begin{aligned} r(W) &= T + r(\Lambda) - \dim(\langle T^*\rangle \cap \langle W_2\rangle) \\ &= T + r(\Lambda) - 0 \\ &= T + r(\Lambda) \end{aligned}$$

If $\langle T^*\rangle \cap \langle W_2\rangle = \{\mathbf{0}_n\}$ then $r(\Lambda)$ is at most $T(B-1)$, otherwise, it leads to the contradiction $r(W_2) > r(X_I)$, thus, if $r(\Lambda) = p = T(B-1)$ then:

$$\begin{aligned} r(W) &= T + T(B-1) \\ &= TB \\ &= r(X_I) \\ &= r(X^*) \end{aligned} \tag{14}$$

but $W_2 \subseteq \langle X_I\rangle$ and $T^* \subseteq \langle X_I\rangle \Rightarrow W \subseteq \langle X_I\rangle = \langle X^*\rangle$; consequently by (14) $\langle W\rangle = \langle X^*\rangle$, thus the models are equivalent as defined in the manuscript.

b. Recall that $W_2 \subseteq \langle X_I\rangle$ and $r(W_2) = r(\Lambda)$ which means that $r(\Lambda) \leq r(X_I) = TB$, thus, if $r(\Lambda) = TB$ then $\langle W_2\rangle = \langle X_I\rangle$ which implies that $\langle T^*\rangle \cap \langle W_2\rangle = \langle T^*\rangle$ and therefore

$$\begin{aligned} r(W) &= T + TB - T \\ &= TB \\ &= r(X^*) \end{aligned}$$

and the equivalence follows.

***iv)*** Notice that $r(W_2) = r(\Lambda) \leq \min\{TB, p\}$, but $r(W_2) \leq r(X_I) = TB$, thus the proof is the same of part *iv)* of Theorem 1.

**Q.E.D**

**Corollary 1**.

**1.** *Small sample precision*. If condition CH1 holds and $p \leq B - 1$, then

$$SSE_X(n - T - p) \leq SSE_W(n - T - B + 1)$$

is a necessary and sufficient condition for: $\hat{\sigma}_X^2 \leq \hat{\sigma}_W^2$. This holds for the scenario discussed in the second part of Theorem 2 if $\Lambda_1$ is null.

**2.** *Large sample behavior of precision*. Consider $n \to \infty$.

2.1. Under condition CH1 or $W_2 \subset \langle X\rangle$; suppose that models 1 and 2 are not equivalent (see part iii of Theorem 1). If model 2 is valid, which implies that $\hat{\sigma}_W^2$ is consistent, then, asymptotically, the models yield the same precision.

2.2. Under condition CH1 or $W_2 \subset \langle X \rangle$, if model 1 holds, but model 2 does not, for $n$ large enough, $\hat{\sigma}_X^2 < \hat{\sigma}_W^2$, disregarding if $\hat{\sigma}_W^2$ is a convergent sequence or not.
2.3. Parts 2.1 and 2.2 hold when comparing models 2 and 3, with model 3 playing the role of model 1 in 2.1 and 2.2.
**3.** If condition CH1 holds, or $W_2 \subset \langle X \rangle$, then $\hat{\sigma}_{ML_X}^2 \leq \hat{\sigma}_{ML_W}^2$, also, if condition CH2 holds, then $\hat{\sigma}_{ML_{X^*}}^2 \leq \hat{\sigma}_{ML_W}^2$.

**Proof**.
**1.** From Theorem 1, it follows that, if $p \leq B - 1$

$$\begin{aligned} & r(W) \leq T + p \\ \Leftrightarrow\ & n - r(W) \geq n - T - p \\ \Leftrightarrow\ & \frac{1}{n - r(W)} \leq \frac{1}{n - T - p} \end{aligned} \tag{15}$$

now, recall that $r(W) \leq r(X) = T + B - 1$, then

$$\begin{aligned} \hat{\sigma}_X^2 \leq \hat{\sigma}_W^2 \Leftrightarrow\ & \frac{SSE_X}{n - r(X)} \leq \frac{SSE_W}{n - r(W)}\ (by\ definition) \\ \Leftrightarrow\ & \frac{SSE_X}{n - T - B + 1} \leq \frac{SSE_W}{n - r(W)} \end{aligned} \tag{16}$$

by (15) and the fact that $SSE_W \geq 0$ it follows that

$$\frac{SSE_W}{n - r(W)} \leq \frac{SSE_W}{n - T - p}$$

by this inequality and (16):

$$\begin{aligned} \hat{\sigma}_X^2 \leq \hat{\sigma}_W^2 \Leftrightarrow\ & \frac{SSE_X}{n - T - B + 1} \leq \frac{SSE_W}{n - T - P} \\ \Leftrightarrow\ & SSE_X(n - T - P) \leq SSE_W(n - T - B + 1) \end{aligned}$$

the last step follows because $n - T - p \in \mathbb{Z}_+$, $n - T - B + 1 \in \mathbb{Z}_+$.

Besides, recall that if $W_2 \subset \langle X \rangle$ then it can be written as:

$$X\Lambda = [T^* \vdots X_2] \begin{bmatrix} \Lambda_1 \\ \cdots \\ \Lambda_2 \end{bmatrix} = T^*\Lambda_1 + X_2\Lambda_2$$

hence, if $\Lambda_1$ is null, then $W_2 = X_2\Lambda_2$ which means that we are in the same setup of theorem 1 and; therefore, the result holds for the second part of theorem 2.
**2.** In order to make the dependence of error sums of squares on the sample size explicit, let us introduce a slight change in notation, $SSE_{X_n}, SSE_{W_n}, SSE_{X_n^*}$, for models 1, 2 and 3, respectively. Similarly, let the unbiased estimates be denoted as $\hat{\sigma}_{X_n}^2, \hat{\sigma}_{W_n}^2, \hat{\sigma}_{X_n^*}^2$ and let:

$$\delta_n := \frac{SSE_{X_n}}{SSE_{W_n}} > 0$$

$$\lambda_n := \frac{n - r(W)}{n - r(X)} \geq 1.$$

2.1. We follow the ideas discussed in Christensen (2011) pg 52-59. Under condition CH1 or $W_2 \subset \langle X \rangle$, model 2 is a reduced version of model 1 because $\langle W \rangle \preccurlyeq \langle X \rangle$, which implies that if model 2 is valid, then model 1 is valid as well (Christensen, 2011). Thus, if these models are not equivalent, since both hold, $\hat{\sigma}^2_{X_n}$ and $\hat{\sigma}^2_{W_n}$ converge to $\sigma^2$, so they are asymptotically equivalent in terms of precision.

2.2. Under condition CH1 or $W_2 \subset \langle X \rangle$, by theorems 1 and 2, respectively, $0 \leq \delta_n \leq 1 \;\forall\; n$. Furthermore, since the sequence $\lambda_n$ converges to 1 so does the sequence $\lambda_n^{-1}$. If model 2 does not hold and the sequence $\hat{\sigma}^2_{W_n}$ converges, then $\hat{\sigma}^2_{W_n} \xrightarrow{n\to\infty} \sigma_*^2 > 0, \sigma_*^2 \neq \sigma^2$, but

$$
\begin{aligned}
1 \geq \delta_n &= \frac{SSE_{X_n}}{SSE_{W_n}} \\
&= \left(\frac{n - r(X)}{n - r(W)}\right)\left(\frac{\hat{\sigma}^2_{X_n}}{\hat{\sigma}^2_{W_n}}\right) \qquad (17)
\end{aligned}
$$

thus

$$
\begin{aligned}
1 &\geq \lim_{n\to\infty} \delta_n \\
&= \lim_{n\to\infty}\left(\frac{n - r(X)}{n - r(W)}\right)\left(\frac{\hat{\sigma}^2_{X_n}}{\hat{\sigma}^2_{W_n}}\right) \\
&= \lim_{n\to\infty}\left(\frac{\hat{\sigma}^2_{X_n}}{\hat{\sigma}^2_{W_n}}\right) \\
&= \frac{\sigma^2}{\sigma_*^2}
\end{aligned}
$$

then, necessarily $\sigma^2 \leq \sigma_*^2$. On the other hand, if the sequence $\hat{\sigma}^2_{W_n}$ does not converge, then $\delta_n$ does not converge either, but $0 \leq \delta_n \leq 1$ and $\lambda_n \xrightarrow{n\to\infty} 1$; therefore, in the limit $\lambda_n \delta_n \leq 1$; furthermore, by expression 17

$$
\hat{\sigma}^2_{X_n} \leq \hat{\sigma}^2_{W_n} \frac{n - r(W)}{n - r(X)} = \hat{\sigma}^2_{W_n} \lambda_n
$$

consequently, for $n$ large enough, $\hat{\sigma}^2_{X_n} \leq \hat{\sigma}^2_{W_n}$, albeit the non-convergence of $\hat{\sigma}^2_{W_n}$ for otherwise, the asymptotic condition $\lambda_n \delta_n \leq 1$ would not be satisfied.

2.3. The proof of this part is essentially the same of 2.1 and 2.2, replacing $\hat{\sigma}^2_{X_n}$ by $\hat{\sigma}^2_{X_n^*}$ and noticing that under the conditions of Theorem 3, $\langle W \rangle \preccurlyeq \langle X^* \rangle$.

**3.** According to Theorem 1, under condition CH1 or $W_2 \subset \langle X \rangle$: $SSE_X \leq SSE_W$; from this inequality and the fact that $n \in \mathbb{Z}_+$ it follows immediately that:

$$\frac{SSE_{X_n}}{n} \leq \frac{SSE_{W_n}}{n} \Leftrightarrow \hat{\sigma}^2_{ML_X} \leq \hat{\sigma}^2_{ML_W}$$

Morevover, making the stated replacements and noticing that the sequence $\frac{n-r(X^*)}{n-r(W)}$ also converges to 1, and that under condition CH2: $\delta_n^* \coloneqq \frac{SSE_{X_n^*}}{SSE_{W_n}} \leq 1$, the proof is completely analogous, thus $\hat{\sigma}^2_{ML_{X^*}} \leq \hat{\sigma}^2_{ML_W}$.

**Q.E.D**

**Corollary 2**. Assume that the conditions stated in Theorems 1, 2 or 3 to guarantee that any estimable function in the one-way Anova is also estimable in models 2 or 3. Then, these functions are estimated with at least the same precision under model 1 than under model 2, and under model 3 than under model 2, provided that parts 1 or 2 of corollary 1 are fulfilled.

**Proof**. Following Christensen (2011 pg. 216-219), since $X_2$ is of full column rank, any estimable function under the one-way anova is also estimable under model 1 (the two-way anova without interaction). Furthermore, under conditions stated in parts $iv$ of theorem 1, or $v$ if theorem 2, it follows that any estimable function in the one-way Anova is also estimable in model 2, similarly, under the conditions of part $iv$ of theorem 3, any estimable function in the one-way Anova is also estimable in model 3. Hence, provided these conditions hold, any linear function of treatment means that is estimable under the one-way Anova model can be estimated in any of the three models considered here. Let $\lambda'\theta_1$ be an estimable function of location parameters associated to treatment means, from linear models theory, the variance of $\lambda'\hat{\theta}_1$ is known to be

$$Var\left[\lambda'\hat{\theta}_1\right] = \sigma^2\lambda'(T^{*\prime}T^*)^-\lambda$$

and, consequently, it can be estimated using an estimator of $\sigma^2$. Now, corollary 1 provides small and large sample conditions leading to $\hat{\sigma}^2_{W_n} \leq \hat{\sigma}^2_{X_n^*}$ or $\hat{\sigma}^2_{W_n} \leq \hat{\sigma}^2_{X_n}$; since $\lambda'(T^{*\prime}T^*)^-\lambda$ is a nonnegative definite quadratic form, the result follows.

**Q.E.D**

**Supplementary file 2.**
**Discrepancy between the true mean of the response variable and its expectations under models 1 and 2 for the case of soil heterogeneity**

Now, let us discuss the relation between the true mean and the expected value of the response variable under models 1 and 2. What is meant by true mean? Blocks are expected to account for a heterogeneity pattern in the soil that is induced by variation in its chemical, physical and biological properties (Behera et al., 2018), sometimes topographical characteristics are also taken into account because topography is known to affect soil properties (Florinsky, 2016). Thus, assuming that $E[Y]$ is a function of a collection of $q$ soil variables seems to be a sound approach. Hence, the true mean is a real-valued function $g(\cdot)$ of $q$ continuous variables $\{w_l\}_{l=1}^{q}$ corresponding to soil properties that are measured at each seeding point as well the means of $T$ treatmens (groups). Even though the true mean of the response variable is not defined as a function of block means, notice that in this problem, the index set $\{i, j, k\}$ is in one-to-one correspondence to the spatial position of each experimental unit, that is, with each seeding point; consequently, the notation $Y_{ijk}$ will be used in model 2 because soil variables may vary from one seeding point to another. Thus, let

$$\mu_{w_ijk} := E\left[Y_{ijk}\right] = g\left(\mu_j, w_{ijk1}, w_{ijk2}, \dots, w_{ijkq}\right)$$

where $Y_{ijk}$ is the response of the $i^{th}$ experimental unit from the $j^{th}$ treatment and $k^{th}$ block, $\mu_{w_ijk}$ is its true mean, $\left\{w_{ijkl}\right\}_{i,j,k,l=1}^{m,T,B,q}$ are observable continuous random variables corresponding to the $l^{th}$ soil property at seeding point $(i, j, k)$, so $r := mB$ is the number of replicates per treatment. The notation $\mu_{w_ijk}$ emphasizes the dependence of the true means on the $q-$dimensional real vector $\boldsymbol{w}_{ijk} := \left(w_{ijk1}, w_{ijk2}, \dots, w_{ijkq}\right)$. Because the relation between the response variable and soil properties is complex and unknown (Thornley and France, 2007), it does not seem appropriate to assume a linear form of $g(\cdot)$ nor any particular functional form. Linear models are being used as an approximation to such a relation; consequently, in this section we discuss the discrepancy between $\mu_{w_ijk}$ and the expected values of $Y_{ijk}$ under models 1 and 2.

Intuitively speaking, when the pattern of spatial distribution of soil composition variables is so complex that values of the covariates within each block-treatment subclass display a high variability, it seems reasonable to expect that discrepancy between the expected value of the response under model 1 and the true mean increases with intra block-treatment subclass variability. This behavior follows because under this model, the expected value of the observations from the same treatment-block combination is the same. Of course, this phenomenon applies for the generalized complete block design, that is, when $m \geq 2$.

**Remark 1**. Notice that when defining $E\left[Y_{ijk}\right]$, $q$ soil variables are mentioned as opposed to $p$. This is done intentionally because the group of soil variables to be measured are defined on the basis of domain-specific knowledge and there is no guarantee that all variables effecting the response are considered; because, for example, not all of them are known or not all of them can be observed due to economic constrains.

Resuming, we will focus on the difference between the true mean (as defined above) and the expectations of $Y_{ijk}$ under model 1 and model 2 as a measure of model adequacy, hereinafter, this parameter will be referred to as MED (mean expected discrepancy). To this end, it will be useful to use scalar notation, thus, model 1 is as follows:

$$Y_{ijk} = \mu + \tau_j + B_k + \varepsilon_{1ijk}$$

where $Y_{ijk}$ is the response of the $i^{th}$ experimental unit from the $j^{th}$ treatment and $k^{th}$ block, $\mu$ is the intercept, $\tau_j$ the effect of the $j^{th}$ treatment, $B_k$ the effect of the $k^{th}$ block and $\varepsilon_{1ijk}$ is the error, errors are assumed to follow IID $N(0, \sigma^2)$ distributions, $i = 1,2, \dots, m; j = 1,2, \dots, T; k = 1,2, \dots, B,$ . Similarly, model 2 is

$$Y_{ijk} = \mu + \tau_j + \sum_{l=1}^{p} \beta_l w_{ijkl} + \varepsilon_{2ijk}$$

where $\beta_l$ is the regression coefficient of the $l^{th}$ covariate, $w_{ijkl}$ is the observed value of the $l^{th}$ covariate for the $i^{th}$ experimental unit from the $j^{th}$ treatment and $k^{th}$ block, and $\varepsilon_{2ijk}$ the error, the remaining terms and the probabilistic assumptions for errors are the same as in model 1.

Under this parametrization $\tau_j := \mu_j - \mu$; $B_k := \mu_k^* - \mu$ where $\mu_j$ is the mean of the $j^{th}$ treatment and $\mu_k^*$ the mean of the $k^{th}$ block. Recall that $\mu_{w_ijk}$ is the true mean of the response variable of the $i^{th}$ experimental unit assigned to the $j^{th}$ treatment in the $k^{th}$ block with soil characteristics $\boldsymbol{w}_{ijk} \in \mathbb{R}^q$. Now, define

$$\mu_{ijk}^B := \mu + \tau_j + B_k$$
$$\mu_{ijk}^A := \mu + \tau_j + \sum_{l=1}^{p} \beta_l w_{ijkl}.$$

These are the expected values of the response variable under models 1 and 2, respectively. So, MED's under models 1 and 2 are:

$$MED_{B_{ijk}} := \mu_{ijk}^B - \mu_{w_ijk}$$
$$MED_{A_{ijk}} := \mu_{ijk}^A - \mu_{w_ijk}.$$

These are the entries of $n-$dimensional vectors $MED_B$ and $MED_A$ whose squared $L_2$ norms $\|MED_B\|_2^2$ and $\|MED_A\|_2^2$ can be used as criteria for model comparison.

The first issue to be studied is related to the impact of $B_k$ on $MED_{B_{ijk}}$. If blocking excels in controlling for SH, some block means should be different, which implies that the corresponding block effects are different from zero, then the following question arises: What happens when all block effects are null? Recall that

$$\begin{aligned} MED_{B_{ijk}} &= \mu + \tau_j + B_k - \mu_{w_ijk} \\ &= \mu_j + B_k - \mu_{w_ijk} \end{aligned}$$

thus, if $B_k = 0$ then $MED_{B_{ijk}} = \mu_j - \mu_{w_ijk}$. Let $\mu_{jk} \coloneqq \mu_j + B_k = \mu_j + \mu_k^* - \mu$, then, squared MED under both scenarios (null and non-null block effects) satisfies

$$\begin{aligned}(\mu_{jk} - \mu_{w_ijk})^2 \leq (\mu_j - \mu_{w_ijk})^2 &\Leftrightarrow \frac{\mu_{jk} + \mu_j}{2} \leq \mu_{w_ijk} \\ &\Leftrightarrow \mu_j + \frac{B_k}{2} \leq \mu_{w_ijk} \end{aligned} \tag{3}$$

Similarly, $(\mu_{jk} - \mu_{w_ijk})^2 > (\mu_j - \mu_{w_ijk})^2 \Leftrightarrow \mu_j + \frac{B_k}{2} > \mu_{w_ijk}$. Therefore, any order relation between MED's with null and non-null block effects is possible, consequently, null block effects do not necessarily increase MED, it depends on the relation between treatment means, block effects and the true mean of the response variable as shown in Equation 3. Now, let us consider $MED^2_{B_{ijk}}$ as a function of $B_k$, then, it is easy to see that it is minimized when $B_k = \mu_{w_ijk} - \mu_j$, thus, squared MED increases as $B_k$ departs from $\mu_{w_ijk} - \mu_j$, that is, it is an increasing function of the distance between $B_k$ and $\mu_{w_ijk} - \mu_j$. Regarding model 2, let $C_{ijk} \coloneqq \sum_{l=1}^{p} \beta_l w_{ijkl}$, thus, $MED_{A_{ijk}}$ can be written as:

$$MED_{A_{ijk}} = \mu + \tau_j + C_{ijk} - \mu_{w_ijk}.$$

Note that the only difference in the MED's for models 1 and 2 is that in $MED_{A_{ijk}}$, $B_k$ is replaced by $C_{ijk}$; as a consequence, it is expected that the relation between squared MED's depends on the difference of these terms. It is easily shown that:

$$\begin{aligned} &MED^2_{A_{ijk}} \leq MED^2_{B_{ijk}} \\ \Leftrightarrow &(C_{ijk} - B_k)\left(MED_{A_{ijk}} + MED_{B_{ijk}}\right) \leq 0. \end{aligned}$$

Consequently, the relation not only depends on the difference between $C_{ijk}$ and $B_k$, but also on the sign of the sum of the MED's. Thus, different signs of the terms $C_{ijk} - B_k$ and $MED_{A_{ijk}} + MED_{B_{ijk}}$ is a sufficient and necessary condition to get a smaller MED with model 2. Again, it is worth noticing that no particular form was given to $\mu_{w_ijk}$ because it provides a more general framework.

The parameter defined as MED (mean expected discrepancy) can be seen as an expected bias for the response variable, that is, the expected difference between its true mean and its mean computed according to each one of the two models. Linear models are used because of their relative simplicity and because they have proven to be very useful in several fields; however, the expected value of the response variable may not be a linear function of model parameters. Thus, in this study the true mean was seen as a function of treatment means and soil variables at each seeding point, and no particular functional form was assumed because it is not necessary, it suffices to assume that it is a finite real number. On the other hand, we have the first moments under model 1 and model 2, that is, the expected value of the response variable computed under each one of them and MED was defined as the difference of these values and the true mean. It was found that adding block effects to the one-way Anova model does not necessarily reduces MED; in fact, it increases as block effects depart from the difference between the true mean and treatment means. Also, the order relationship between $MED^2_{A_{ijk}}$ and $MED^2_{B_{ijk}}$ depends not only on the difference of the aggregate effect of covariates ($C_{ijk}$) and block effects ($B_k$), but also on the sign of $MED_{Aijk} + MED_{Bijk}$. Besides, notice that the conclusions drawn for MED also hold is we replace the true mean $\mu_{w_ijk}$ by the observed response $y_{ijk}$ since it is a realized value of $Y_{ijk}$ and, consequently, it is a nonrandom quantity.

## Appendix 2: Details on the simulation study

***Simulation procedure***

As stated in the manuscript data were simulated in two steps, the first one involved the simulation of soil characteristics and the second one the simulation of records as functions of three components: treatment means, a function of soil characteristics, and errors. Soil variables were simulated as described in the manuscript. Using the premise that many variables of agronomical interest such as biomass production depend on soil properties (Thornley and France, 2007), this simulation modeled the response variable as a function of a collection of continuous variables representing soil characteristics, the treatment means and a random error. Therefore, once the four soil variables were simulated, the expected value of the response was obtained as follows:

$$\mu_{ijk} = g\left(\mu_{.j.}, W_{ijk}\right) = \mu_{.j.} + g_2^{\gamma}\left(W_{ijk}\right)$$

where $\mu_{ijk}$ is the expected value of the $i^{th}$ experimental unit from the $j^{th}$ treatment and $k^{th}$ block, $\mu_{.j.}$ is the mean of the $j^{th}$ treatment, $W_{ijk} \coloneqq \left(w_{1ijk}, w_{2ijk}, w_{3ijk}, w_{4ijk}\right)$ is the vector containing the simulated values of the four soil variables, $g_2^{\gamma}: \mathbb{R}^4 \to \mathbb{R}$ is a continuous function of the soil variables which has the form

$$g_2^{\gamma}\left(W_{ijk}\right) = \sum_{l=1}^{4} g_{\gamma}^{(l)}\left(w_{lijk}\right)$$

Two functional forms were assumed for $g_{\gamma}^{(l)}$

$$g_{1_{\gamma}}^{(l)}(x) = \gamma_{l1} e^{\left(-e^{(-\gamma_{l2}x)}\right)}$$

$$g_{2_{\gamma}}^{(l)}(x) = \gamma_{l1}\left(1 - e^{(-\gamma_{\ell 2}x)}\right)$$

$$(\gamma_{l1}, \gamma_{l2}) \in \mathbb{R}_{+}^{2}$$

where $\mathbb{R}_{+}^{2}$ is the set of two-dimensional vectors whose entries are positive real numbers.

As pointed out above, these forms were chosen to emulate the typical behavior of many traits of economic importance as functions of soil nutrients (Thornley and France, 2007); however, in order to introduce some variability, we chose two different functions featuring such a behavior but having different rates of change (i.e., first derivatives). For each one of the four soil variables, a Bernoulli(1/2) random variable was simulated to decide which function was employed; so if the realized value of the Bernoulli variable was equal to zero, $g_{1_{\gamma}}^{(l)}$ was used, while $g_{2_{\gamma}}^{(l)}$ was used otherwise.

On the other hand, in order to simulate scenarios ranging from all treatment means being equal to all being different, the following procedure was adopted. Recall that the total number of treatments was 4. In the first step, $K \in \{1,2,3,4\}$ was drawn from the corresponding discrete uniform distribution (i.e., a discrete uniform distribution with support set {1,2,3,4}). In the second step, if $K \in \{2,3\}$, a random partition of the first 4 positive integers of size $K$ was created, let such a partition be defined as:

$$\wp_{(1,4,K)} := \{\wp_1, \dots, \wp_K\}, K \in \{2,3\}$$

then, the first $K$ elements of the vector $\alpha := (30, 35, 42, 45)$ were used to sample the treatment means from a Gaussian distribution as detailed bellow. The cases $K = 1$ and $K = 4$ correspond to all means being equal and all being different, respectively.

Now, let $F_{W_M}(\cdot)$ be the distribution function corresponding to the Gaussian process described in Zhang (2004) that was used to simulate the soil variables. Then, the following is the simulation procedure.

Set the following entries:

$$(\alpha_1, \alpha_2, \dots, \alpha_T) \in \mathbb{R}^T$$
$$(a_1, a_2), (b_1, b_2), (sh, sc) \in \mathbb{R}_+^2, a_2 > a_1, b_2 > b_1$$
$$m, q, T \in \mathbb{Z}_+$$

Compute

$$n = mTB$$

Build the design matrix for treatment means $M_{n \times T}$ (using the 1-of-k encoding)

Generate

$$W \sim F_{W_M}(\cdot)$$
$$K \sim Unif.Discrete(1, T)$$

If $K \in \{2,3\}$ then create the random partition of size $K$ of the first four positive integers

$$\wp_{(1,4,K)} := \{\wp_1, \dots, \wp_K\}$$

so $\wp_i, i = 1, \dots, K$ are mutually exclusive subsets of the set of the first four positive integers.

If $K = 1$ set

$$\wp_{(1,4,K)} = \{1,2,3,4\}$$

If $K = 4$ set

$$\wp_{(1,4,K)} = \{\{1\}, \{2\}, \{3\}, \{4\}\}$$

Sample

$$\sigma_\mu^2 \sim Gamma(sh, sc)$$

For $i = 1, \dots, K$:

Sample

$$\delta_i^* \overset{ind}{\sim} N(\alpha_i, \sigma_\mu^2)$$

Set

$$\mu_j = \delta_i^* \ \forall\ j \in \wp_i, j = 1,2, \dots, T$$

Build the vector

$$\mu^* := (\mu_1, \mu_2, \dots, \mu_T)'$$

Sample

$$\sigma_\varepsilon^2 \sim Gamma(0.9sh, 0.8sc)$$
$$\varepsilon \sim N(0, \sigma_\varepsilon^2 I_n)$$

for $h = 1,2$ sample

$$\gamma_{1h} \overset{ind}{\sim} Unif.Cont.(a_1, a_2), \gamma_{2h} \overset{ind}{\sim} Unif.Cont.(b_1, b_2)$$

$$\gamma_l := \begin{pmatrix} \gamma_{l1} \\ \gamma_{l2} \end{pmatrix}$$

where $Unif.Cont$ stands for uniform continuous.

For $l = 1,2, \dots, q$

$$Z_l^* \overset{ind}{\sim} Bernoulli\left(\frac{1}{2}\right)$$

$$g_\gamma^{(l)} = \begin{cases} g_{1_{(\gamma_{11},\gamma_{21})}}^{(l)}, if\ Z_l^* = 0 \\ g_{2_{(\gamma_{12},\gamma_{22})}}^{(l)}, Otherwise \end{cases}$$

Create the vector

$$\Lambda := \left\{ \sum_{l=1}^{q} g_\gamma^{(l)} \left( w_{lijk} \right) \right\}_{n\times 1}$$

Finally, compute the simulated response vector as

$$y = M\mu^* + \Lambda + \varepsilon$$

Repeat the process $R$ times.

In this case

$$\alpha = (30, 35, 42, 45)$$
$$q = T = 4, m = 2, B \in \{3,4,6\}$$
$$(a_1, a_2) = (15, 22), (b_1, b_2) = (0.02, 0.1), (sh, sc) = (3, 2)$$

The values of $a_1, a_2, b_1, b_2$ were chosen to have functions with a low curvature, so that models assuming linear trends could be appropriate approximations.

***Accuracy of the test for additive effects of treatments***

The ability of each model to correctly test the null hypothesis of no additive treatment effects (equivalent to equality of treatment means) was computed as follows.

Using the results of the 500 replicates, the following two-way contingency table was built

| Number of true positives (NTP) | Number of false positives (NFP) |
|---|---|
| Number of false negatives (NFN) | Number of true negatives (NTN) |

A true positive occurred when $K \neq 1$ and the null hypothesis was rejected, while a false positive occurred when $K = 1$ and the null hypothesis was rejected. Similarly, a true negative occurred when $K = 1$ and the null hypothesis was not rejected, whereas a false negative occurred when $K \neq 1$ and the null hypothesis was not rejected.

The corresponding accuracy was computed as:

$$Acc = \frac{NTP + NTN}{500}$$

**Supplementary Table 1.** Across-replicates summary of the simulation study in different setups (the two models under different scenarios) for the mean absolute and squared errors of the observed response and estimated pairwise differences between treatment means.

| Setup[1] | Missing data % | RMSEy | MAEy | RMSEE | MAEE |
|---|---|---|---|---|---|
| Model 1 3B | -- | 5.57 | 4.49 | 14.00 | 8.00 |
| Model 1 4B | -- | 5.72 | 4.61 | 13.59 | 7.54 |
| Model 1 6B | -- | 5.80 | 4.64 | 13.41 | 7.11 |
| Model 2 NMC | 0 | 3.91 | 3.14 | 13.55 | 7.04 |
| Model 2 MC | 10 | 4.10 | 3.27 | 13.57 | 7.11 |
| Model 2 MC | 20 | 4.29 | 3.40 | 13.58 | 7.20 |
| Model 2 MC | 30 | 4.49 | 3.55 | 13.60 | 7.27 |
| Model 2 MC | 40 | 4.70 | 3.70 | 13.63 | 7.35 |
| Model 2 MC | 50 | 4.94 | 3.90 | 13.65 | 7.48 |
| Model 2 MC | 60 | 5.12 | 4.05 | 13.69 | 7.54 |

[1]Same conventions as in Table 1.
RMSEy: squared root of mean squared error of the response variable, MAEy: mean absolute error of the response variable, RMSEE: squared root of mean squared error of pairwise treatment differences, MAEE: mean absolute error of pairwise treatment differences.

**Supplementary Table 2.** Relevant results from the analysis of variance for average daily gain at 70 days from the piglets' experiment using model 1 (two-way Anova without interaction) and model 2 (Ancova) with different covariates. Results from equivalent models are shown in bold

| **Model[1]** | **Source** | **DF** | **SS** | **MS** | **F Stat.** |
|---|---|---|---|---|---|
| **Two-way Anova** | **Diet** | **4** | **0.031** | **0.008** | **3.238** |
| | **Error** | **53** | **0.127** | **0.002** | |
| Ancova: Tmin | Diet | 4 | 0.031 | 0.008 | 3.272 |
| | Error | 54 | 0.128 | 0.002 | |
| **Ancova: TMin + MaxDeltaT** | **Diet** | **4** | **0.031** | **0.008** | **3.238** |
| | **Error** | **53** | **0.127** | **0.002** | |
| **Ancova: Tmax + TotPrec** | **Diet** | **4** | **0.031** | **0.008** | **3.238** |
| | **Error** | **53** | **0.127** | **0.002** | |
| **Ancova: TotPrec + MinPR + MinHR** | **Diet** | **4** | **0.031** | **0.008** | **3.238** |
| | **Error** | **53** | **0.127** | **0.002** | |

[1]For the Ancova models, abbreviations after the colon correspond to covariates included in the model. Tmin: minimum temperature, MaxDeltaT: maximum temperatura change, Tmax: máximum temperature, TotPrec: total precipitation, MinPR: minimum precipitation, MinHR: minimum relative humidity.

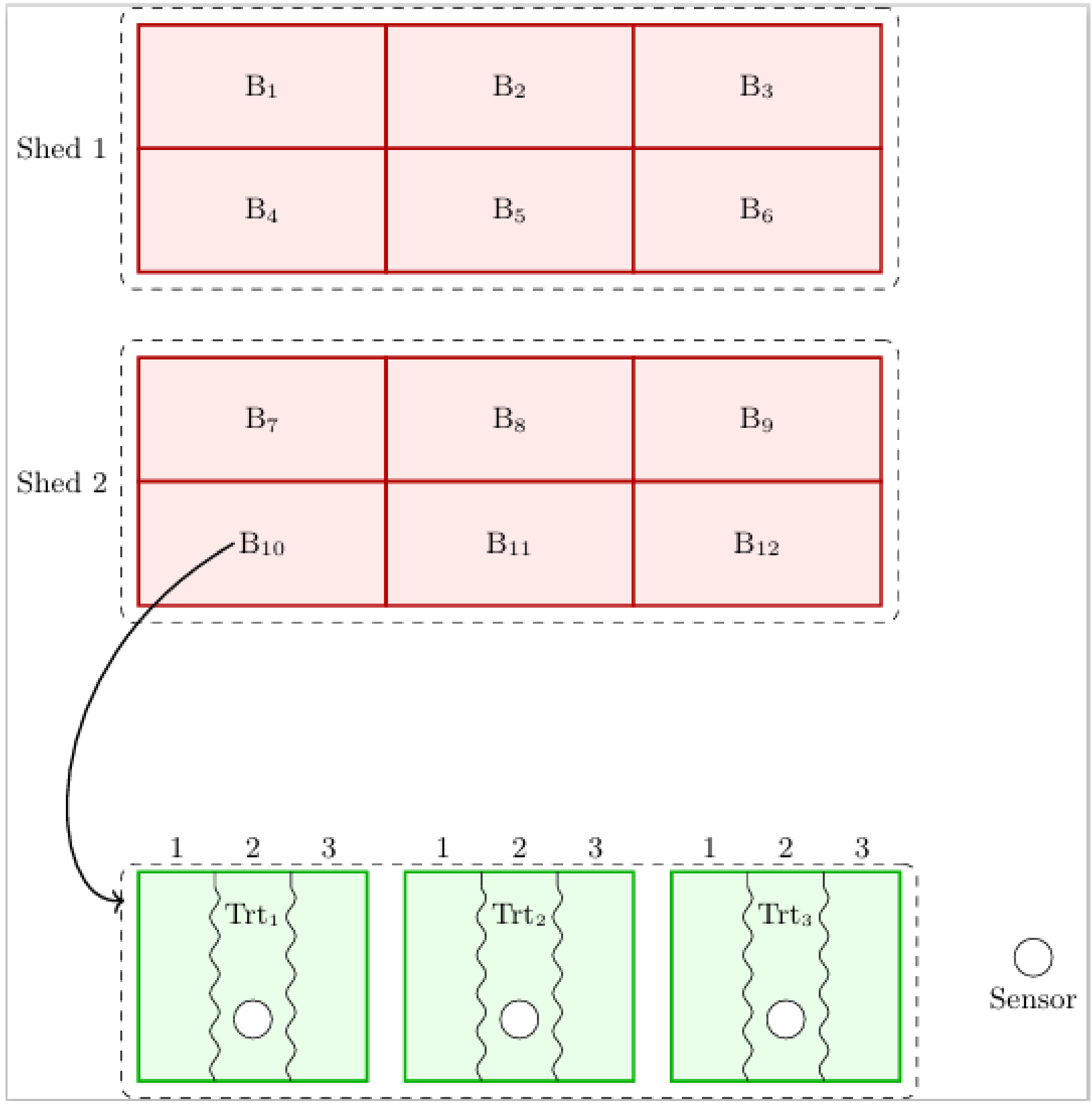

**Supplementary Figure 1**. Arrangement of the layers experiment. It portrays a single row of two sheds and a zoom of a block. There are three treatment groups and a sensor is located on each treatment-block combination, hence, condition CH2 is met.

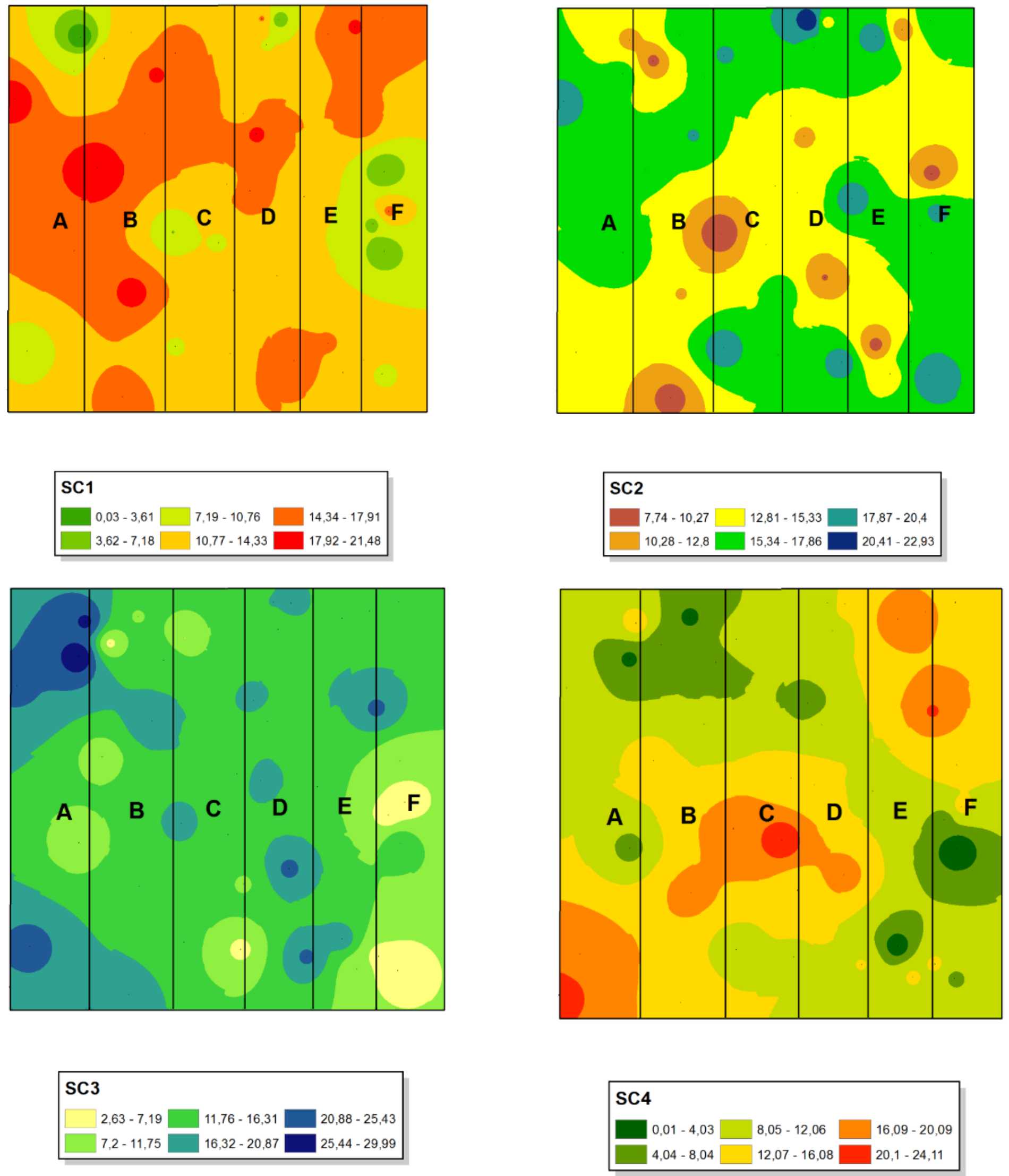

**Supplementary Figure 2**. Spatial distribution of the four simulated soil composition variables (SC1, SC2, SC3, SC4) for replicate 1 under the six blocks scenario. Letters correspond to blocks.